\documentclass[a4paper,fleqn,usenatbib]{mnras}
\usepackage[T1]{fontenc}
\usepackage{ae,aecompl}
\usepackage{amsmath}
\usepackage{amsfonts}
\usepackage{amssymb}
\usepackage{graphicx}
\usepackage{color,xcolor}
\usepackage{tabularx}
    \newcolumntype{L}{>{\raggedright\arraybackslash}X}
\def\mnras{MNRAS}
\def\aap{A$\&$A}
\def\nat{Nature}
\def\apj{ApJ}
\def\apjl{ApJ Letters}
\def\apjs{ApJS}

\def\aj{AJ}

\newcommand{\me}{\, {\rm M}_{\oplus}}
\newcommand{\msun}{\, {\rm M}_{\odot}}
\newcommand{\msunyr}{\, {\rm M}_{\odot}\,{\rm yr^{-1}}}

\newcommand{\au}{\, {\rm au}}
\newcommand{\rgeuv}{\, {r_{\rm g,euv}}}
\newcommand{\rgfuv}{\, {r_{\rm g,fuv}}}

\title[Disc Evolution Pathways]{Dispersal of protoplanetary discs: How stellar properties and the local environment determine the pathway of evolution}
\author[G. A. L. Coleman and T. J. Haworth]{Gavin A. L. Coleman\thanks{Email: gavin.coleman@qmul.ac.uk} and  Thomas J. Haworth\\
Astronomy Unit, School of Physics and Astronomy, Queen Mary University of London, Mile End Road, London, United Kingdom}
\date{Accepted 2022 May 26. Received 2022 May 26; in original form 2022 February 15}
\pubyear{2022}
\begin{document}
\label{firstpage}
\pagerange{\pageref{firstpage}--\pageref{lastpage}}
\maketitle
\begin{abstract}
We study the evolution and final dispersal of protoplanetary discs that evolve under the action of internal and external photoevaporation, and different degrees of viscous transport. We identify five distinct dispersal pathways, which are i) very long lived discs ($>20\,$Myr), ii) inside-out dispersal where internal photoevaporation dominates and opens inner holes, iii) outside-in dispersal where external photoevaporation dominates through disc truncation and two intermediate regimes characterised by lingering material in the inner disc with the outer disc dispersed predominantly by either internal or external photoevaporation.  We determine how the lifetime, relative impact of internal and external winds and clearing pathway varies over a wide, plausible, parameter space of stellar/disc/radiation properties. There are a number of implications, for example in high UV environments because the outer disc lifetime is shorter than the time-scale for clearing the inner disc we do not expect transition discs to be common, which appears to be reflected in the location of transition disc populations towards the Orion Nebular Cluster. Irrespective of environment, we find that ongoing star formation is required to reproduce observed disc fractions as a function of stellar cluster age. This work demonstrates the importance of including both internal and external winds for understanding protoplanetary disc evolution.  
\end{abstract}
\begin{keywords}
 accretion, accretion discs -- protoplanetary discs -- (stars:) circumstellar matter
 \end{keywords}

\section{Introduction}
\label{sec:intro}
Planet formation is likely to be intrinsically linked to the evolution of the circumstellar discs of material around young stars \citep[e.g.][]{2018ApJ...869L..41A, 2018A&A...617A..44K, 2018ApJ...860L..13P, 2019Natur.574..378T}. However, understanding this is an extremely challenging problem. We observe each disc at only a snapshot in its evolution, and any given disc is difficult to age. This is further complicated by the fact that star formation happens over time in stellar clusters \citep[e.g.][]{2014ApJ...787..108G, 2014ApJ...787..109G}. Since discs have a huge radial variation in composition, temperature, density and optical depth, any single observation also only probes a limited region of the disc \citep[see e.g. Figure 1 of][]{2016PASA...33...59S}. 

In addition to the observational challenges, discs evolve via a range of complex mechanisms. For example, their dispersal can proceed by accretion \citep[e.g.][]{1981ARA&A..19..137P, 2016A&A...591L...3M, 2020A&A...639A..58M, 2021A&A...650A.196M}, internal photoevaporative \citep[e.g.][]{Alexander07, Owen12, 2017RSOS....470114E} or magnetically driven winds \citep{1991ApJ...376..214B, 2007prpl.conf..277P, 2009ApJ...691L..49S, 2014prpl.conf..411T, 2016ApJ...818..152B} as well as winds driven by  external irradiation from nearby massive stars \citep[e.g.][]{1994ApJ...436..194O, 2000ApJ...539..258R, 2004ApJ...611..360A, 2005AJ....130.1763S, 2016ApJ...826L..15K,  2019MNRAS.485.3895H, 2021MNRAS.501.3502H}. These processes have been studied somewhat independently both from a theoretical and observational perspective. However, in reality they will be acting in tandem to determine how the disc surface density profile evolves over time, including the time-scale on which parts of the disc disappear from the outside-in (due to external photoevaporation) or from the inside out (due to accretion/internal winds). 

Planetary population synthesis models \citep[see for example][and references therein]{Mordasini18} have included recipes for a large number of disc processes, including viscous accretion using an $\alpha$ disc model \citep{Shak}, internal photoevaporation due to high-energy photons from central stars \citep{Clarke2001,Dullemond,Alexander07,Alexander09,Owen12}, and external photoevaporation due to high-energy photons originating from nearby sources \citep{Matsuyama03}.
However the impacts of these processes on to the disc are tuned to ensure that the resulting disc properties match observed statistics such as disc lifetimes. But, as noted above, disc lifetimes are extremely difficult to estimate. Usually they are determined by comparing disc fractions in different clusters \citep{2005astro.ph.11083H, Ribas15, Richert18}, but within any cluster there will also be a potentially large spread in ages due to ongoing star formation \citep[e.g.][]{2014ApJ...787..108G, 2014ApJ...787..109G}. Furthermore, ``disc fraction'' can mean different things depending on the observation used to measure it if different parts of the disc are dispersed at different rates. For example, Spitzer IR excesses probe the very inner disc ($\sim 1\,$au) which could be affected by local viscous evolution, internal photoevaporation, or inner planets.  ALMA continuum observations probe a region of the dust disc, which is observed to be less radially extended than the gas \citep{2016ApJ...832..110C, 2018ApJ...859...21A, 2017A&A...605A..16F, 2019A&A...629A..79T}, and expected to be dynamically decoupled from the gas \citep[e.g.][]{2012A&A...539A.148B}. ALMA line observations of the gas disc (e.g. in CO) probe out to large radii but may be influenced by processes such as chemical evolution \citep[e.g.][]{2017ApJ...841...39Y, 2018A&A...613A..14E,2021ApJS..257....8A,  2021PhR...893....1O}. Overall this makes calibration of disc dispersal processes very difficult, driving a need to better understand them theoretically since they have implications for interpreting disc observations and for models studying the resulting  planet populations.

In this paper we study the evolution and dispersal of viscously evolving discs undergoing combined internal and external photoevaporation. Each process is explored using a parameter space motivated by observations and models of those specific properties (e.g. mass loss rates) rather than the properties and outcomes of the discs themselves. This is distinct from \cite{Hasegawa22} who combined internal MHD winds and external photoevaporation in disc models in a semi-analytic way to ask if one could diagnose dispersal processes observationally using the ratio of wind mass loss rate to accretion rate, rather than following the disc evolution and dispersal. Our goal is to determine the various pathways by which discs evolve and are ultimately dispersed, and how this depends on both the host star/disc properties and star forming environment. 

This paper is organised as follows.
Section \ref{sec:physical_model} outlines the disc evolution and photoevaporation models as well as the simulation parameters.
In sect. \ref{sec:results} we present the evolution pathways that arise from simulations around a single subset of the disc parameter space.
We explore the impact of varying viscosity and stellar mass in sect. \ref{sec:main_plots}.
In sect. \ref{sec:discussion} we discuss the outcomes of the models in comparison with numerous features that have arisen from protoplanetary disc observations, before we draw our conclusions in sect. \ref{sec:conclusions}.

\section{Physical Model and Parameters}
\label{sec:physical_model}

Protoplanetary discs lose mass by accretion onto the central star and through photoevaporative winds launched from the disc surface layers.
To account for these processes we adopt a 1D viscous disc model similar to that used in previous works \citep{ColemanNelson14,ColemanNelson16,Coleman21} where the equilibrium temperature is calculated by balancing irradiation heating from the central star, background heating from the residual molecular cloud, viscous heating and blackbody cooling.
We inititialise the disc surface density, $\Sigma$, by assuming a constant power law gas disc
\begin{equation}
    \Sigma(r) = \Sigma_0 \left(\frac{r}{1\au}\right)^{-1}
\end{equation}
where $\Sigma_0$ is the gas surface density at $1\au$ depending on the initial sizes and masses of the discs considered. This slope is chosen as evolving protoplanetary discs typically reach steady state solutions around -1.
The surface density is then evolved by solving the standard diffusion equation
\begin{equation}
    \dot{\Sigma}(r)=\dfrac{1}{r}\left[3r^{1/2}\dfrac{d}{dr}\left(\nu\Sigma r^{1/2}\right)\right]-\dot{\Sigma}_{\rm PE}(r)
\end{equation}
where $\nu=\alpha H^2\Omega$ is the disc viscosity with viscous parameter $\alpha$ \citep{Shak}, $H$ being the disc scale height, $\Omega$ the Keplerian frequency, and $\dot{\Sigma}_{\rm PE}(r)$ is the rate of change in surface density due to photoevaporative winds.
We include both the extreme ultraviolet (EUV) and X-ray internal photoevaporative winds from the central star (detailed in section \ref{sec:internalPhoto}) as well as winds launched from the outer disc by far ultraviolet (FUV) radiation emanating from nearby massive stars (e.g. O-type stars, see section \ref{sec:externalPhoto}).
We assume (and validate in section \ref{sec:photoCompare}) that the photoevaporative mass loss rate at any radius in the disc is the maximum of the EUV, FUV, and X-ray driven rates:
\begin{equation}
    \dot{\Sigma}_{\rm PE}(r) ={\rm max}\left(\dot{\Sigma}_{\rm I,EUV}(r),\dot{\Sigma}_{\rm I,X}(r),\dot{\Sigma}_{\rm E,FUV}(r)\right)
\end{equation}
where the subscripts I and E refer to contributions from internal and external photoevaporation.
The interplay between internal and external photoevaporative winds in dynamical simulations is currently completely unexplored, with only the internal and external considered separately to date. Internal and external photoevaporation do operate on different regions of the disc, so in the absence of simulations that suggest there is a significant interplay we adopt the assumption that they are separable, and that the photoevaporative winds are driven by whichever mechanism is stronger.

\subsection{Internal Photoevaporation}
\label{sec:internalPhoto}
The absorption of high energy radiation from the host star by the disc can heat the gas above the local escape velocity, and hence drive internal photoevaporative winds. EUV irradiation creates a layer of ionised hydrogen with temperature $\sim$10$^4$~K \citep{Clarke2001}, however X-rays penetrate deeper into the disc and are still capable of heating up to around $\sim$10$^4$~K \citep{2010MNRAS.401.1415O} so for low mass stars are expected to generally dominate over the EUV for setting the mass loss rate. FUV radiation penetrates deeper still, creating a neutral layer of dissociated hydrogen with temperature of roughly 1000K \citep{Matsuyama03}. The overall interplay between the EUV, FUV and X-rays is a matter of ongoing debate.  \cite{Owen12} find that including the FUV heating simply causes the flow beneath the sonic surface to adjust, but otherwise retains the same mass loss rate. However others suggest a more dominant role of the FUV \citep[e.g.][]{2009ApJ...705.1237G,2015ApJ...804...29G}. Recent models including all three fields suggest a more complicated interplay \citep[e.g.][]{2017ApJ...847...11W, 2018ApJ...865...75N}. 

We are not here to weigh in on that debate, but opt to incorporate both EUV and X-ray radiation from the central star, since prescriptions for the FUV and X-ray driven dominated mass loss are similar and for the low stellar masses $\leq1\,$M$_\odot$ they hold the X-ray driven mass loss rates as being stronger. 

\subsubsection{EUV}
Our internal EUV photoevaporation prescription follows \citet{Dullemond} where the surface density is depleted as
\begin{equation}
\label{eq:sig_dot_euv}
\dot{\Sigma}_{\rm I,EUV}(r) = 1.16\times10^{-11}G_{\rm fact}\sqrt{f_{41}}\left(\dfrac{1}{\rgeuv}\right)^{3/2}
\left(\dfrac{M_{\bigodot}}{\au^2 \, {\rm yr}}\right)
\end{equation}
where $G_{\rm fact}$ is a scaling factor defined as
\begin{equation}
G_{\rm fact} = \left\{ \begin{array}{ll}
\left(\dfrac{\rgeuv}{r}\right)^2 e^{\frac{1}{2}\left(1-\dfrac{\rgeuv}{r}\right)} 
& r\le \rgeuv, \\
\\
\left(\dfrac{\rgeuv}{r}\right)^{5/2} & r>\rgeuv.
\end{array} \right.
\end{equation}
Here, $\rgeuv$ is the characteristic radius beyond which gas becomes unbound from the system as a result of the EUV radiation, which is set to $10\au$ for Solar-mass stars, and $f_{41}$ is the rate at which ionising photons are emitted by the central star in units of $10^{41}$ s$^{-1}$. This ionising flux is dominated by the prolonged excess chromospheric activity \citep{Alexander05}, that can persist for extended periods of time ($\geq 10^{6.5}$--$10^{7}$ yrs), with typical values $f_{41}=1-100$ that are relatively insensitive to stellar mass for T Tauri stars \citep[see for example][and references therein]{Dullemond}.    

\begin{table*}
    \centering
    \begin{tabular}{ccccccccc}
    \hline
        $M_*$ ($\msun$) & a & b & c & d & e & f & g\\
        \hline
        0.1 & -3.8337 & 22.91 & -55.1282 & 67.8919 & -45.0138 & 16.2977 & -3.5426\\
        0.3 & -1.3206 & 13.0475 & -53.699 & 117.6027 & -144.3769 & 94.7854 & -26.7363\\
        1.0 & 6.3587 & 6.3587 & -26.1445 & 56.4477 & -67.7403 & 43.9212 & -13.2316\\
        \hline
    \end{tabular}
    \caption{Parameters for the surface density profiles in equations \ref{eq:sig_dot_xray} and \ref{eq:m_dot_r_xray} following \citet{Ercolano21}}
    \label{tab:xray_param}
\end{table*}

\subsubsection{X-Ray}

Internal X-ray driven photoevaporation is important for T Tauri stars because it penetrates a deeper column than the EUV and hence can be key for determining the mass loss rates \citep{2010MNRAS.401.1415O, 2011MNRAS.412...13O, Owen12}. The radiation hydrodynamic models of \cite{Owen12} used pre-computed X-ray driven temperatures as a function of the ionisation parameter ($\xi = L_X / n /r^2$) wherever the column to the central star is less than $10^{22}$cm$^{-2}$ (and hence optically thin). This approach has since been updated with a series of column-dependent temperature prescriptions \citep{Picogna19,Ercolano21,Picogna21}.  

We follow \citet{Picogna21} who further build on the work of \cite{Picogna19} and \cite{Ercolano21} to include the dependence on stellar mass \citep[though see also][]{2021ApJ...910...51K} finding that the mass loss profile from internal X-ray irradiation is approximated by
\begin{equation}
\label{eq:sig_dot_xray}
\begin{split}
\dot{\Sigma}_{\rm I,X}(r)=&\ln{(10)}\left(\dfrac{6a\ln(r)^5}{r\ln(10)^6}+\dfrac{5b\ln(r)^4}{r\ln(10)^5}+\dfrac{4c\ln(r)^3}{r\ln(10)^4}\right.\\
&\left.+\dfrac{3d\ln(r)^2}{r\ln(10)^3}+\dfrac{2e\ln(r)}{r\ln(10)^2}+\dfrac{f}{r\ln(10)}\right)\\
&\times\dfrac{\dot{M}_{\rm X}(r)}{2\pi r} \dfrac{\msun}{\au^2 {\rm yr}}
\end{split}
\end{equation}
where
\begin{equation}
\label{eq:m_dot_r_xray}
    \dfrac{\dot{M}_{\rm X}(r)}{\dot{M}_{\rm X}(L_{X})} = 10^{a\log r^6+b\log r^5+c\log r^4+d\log r^3+e\log r^2+f\log r+g}
\end{equation}
where the parameters $a$--$g$ are given in Table \ref{tab:xray_param} for the different stellar masses studied here \citep{Ercolano21}.
Following \cite{Picogna19} the integrated mass-loss rate, dependant on the stellar X-ray luminosity, is given as
\begin{equation}
    \log\left[\dot{M}_{X}(L_X)\right] = A_{\rm L}\exp\left[\dfrac{(\ln(\log(L_X))-B_{\rm L})^2}{C_{\rm L}}\right]+D_{\rm L},
\end{equation}
in $\msunyr$, with $A_{\rm L} = -2.7326$, $B_{\rm L} = 3.3307$, $C_{\rm L} = -2.9868\times 10^{-3}$, and $D_{\rm L} = -7.258$.

Near the end of disc lifetimes, a hole can form in the inner regions, allowing the radiation from the central star to launch a wind directly from the full inner edge of the disc, rather than just the surface layers.
This increase in photoevaporation rate can then quickly disperse the disc from the inside-out on short time-scales ($\sim$0.1--0.2 Myr).
We continue to follow \citet{Picogna19} and define the mass loss profile when an inner hole has formed as being equal to
\begin{equation}
\label{eq:sig_dot_hole}
\dot{\Sigma}_{\rm I,X}(r)=a_1 b_1^x x^{c_1-1}(x\log{b_1}+c_1)\dfrac{1.12\dot{M}_{X}(L_X)}{2\pi r}\dfrac{\msun}{\au^2{\rm yr}}
\end{equation}
where $x=(r-r_{\rm gap})$, $r_{\rm gap}$ is the size of the inner hole, $a_1=0.11843$, $b_1=0.99695$ and $c_1=0.48835$. Note that in the fits from \citet{Picogna21} there is currently no stellar mass dependent prescription of internal photoevaporation when the inner hole is opened\footnote{However, in previous fits for X-ray internal photoevaporation, \citet{Owen12} did include a stellar mass dependent fit for mass loss rates once an inner hole has been opened, yielding mass loss rates up to an order of magnitude lower for the lowest mass stars considered here. Since the hole determined photoevaporation rates only occur at the end of the disc lifetimes, and do not affect the discs evolutionary pathway, we choose to use the \citet{Picogna21} for consistency in calculating the X-ray photoevaporation rates.}, though once the hole does open the remaining disc lifetime is short so we do not deem this to be significant. 
We define the disc as having opened an inner hole when the radial column density in the inner disc region drops below $10^{22} {\rm cm^{-2}}$ \citep{Owen12}. 

\subsection{External Photoevaporation}
\label{sec:externalPhoto}

In addition to internal winds driven by irradiation from the host star, winds can also be driven from the outer regions of discs by irradiation from external sources. Massive stars dominate the production of UV photons in stellar clusters and hence dominate the external photoevaporation of discs. External photoevaporation has been shown to play an important role in setting the evolution of the  disc mass \citep{2014ApJ...784...82M, 2017AJ....153..240A}, radius \citep{2018ApJ...860...77E} and lifetime \citep{2016arXiv160501773G, 2019MNRAS.490.5678C, 2020MNRAS.492.1279S, 2020MNRAS.491..903W} even in weak UV environments \citep{2017MNRAS.468L.108H}.
Here we include the effects of external photoevaporation due to far-ultraviolet (FUV) radiation emanating from massive stars in the vicinity of the discs following \cite{Matsuyama03}.
In this, a wind is driven outside of the gravitational radius where the sound speed in the heated layer is $T\sim$1000~K, denoted $\rgfuv$.
The gas surface density is depleted as
\begin{equation}
\label{eq:sig_dot_external}
\dot{\Sigma}_{\rm E,FUV}(r) = G_{\rm sm} \dfrac{\dot{M}_{\rm pe,ext}(r_{\rm d})}{\pi(r_{\rm max}^2-\beta^2 \rgfuv^2)}.
\end{equation}
where $\beta = 0.14$ \citep[similar to][]{AlexanderPascucci12} gives the effective gravitational radius that external photoevaporation operates beyond (for example due to pressure gradients in the disc that help drive a wind interior to the basic gravitational radius), $r_{\rm max}$ is the outer disc disc radius calculated where $\Sigma(r)>10^{-4} {\rm gcm}^{-2}$ and $\dot{M}_{\rm pe,ext}(r_{\rm d})$ is the mass-loss rate for a disc of size $r_{\rm d}$.
The factor $G_{\rm sm}$ is a smoothing function located around the effective gravitational radius to adequately determine the mass loss in the inner regions of the disc where external photoevaporation is ineffective,
\begin{equation}
\label{eq:smooth}
    G_{\rm sm} = 1 - \left(1+\left(\dfrac{r}{\beta\rgfuv}\right)^{20}\right)^{-1}.
\end{equation}

To explore the effect of different external photoevaporative mass-loss rates that would be expected in different UV environments we set the mass-loss rate at 100 $\au$, $\dot{M}_{\rm pe, ext, 0}$ as a free parameter with the actual mass loss rate scaling with disc radius as 
\begin{equation}
    \dot{M}_{\rm pe, ext}(r_{\rm d})= \dot{M}_{\rm pe, ext, 0} \left(\frac{r_{\rm d}}{100\au}\right).
    \label{equn:evap}
\end{equation}
As discs are truncated by the external photoevaporative wind, the mass loss rate from external photoevaporation falls, until the disc size reaches the effective gravitational radius and it is entirely shut off.

The above approach to external photoevaporation has the advantage that it is simple to implement and computationally cheap. A  disadvantage is that it is based on a mass loss rate normalisation at 100\,au which does not obviously correlate with any given external FUV field strength. We compared with results from the {\textsc{fried}} grid of external photoevaporative mass loss rates \citep{2018MNRAS.481..452H} and from this assume a roughly linear scaling of the mass loss rate normalization with FUV field strength, with $10^{-7}\msunyr$ corresponding to $10^4$\,G$_0$ for a solar mass star (for lower mass stars the same mass loss rate is achieved for slightly weaker UV field strengths). We will later employ this approximation when placing known transition discs, which we define as discs with SEDs indicating signatures of gaps within the dust and/or gas discs, on plots of our parameter space based on their anticipated ambient FUV environment.
It is also important to note that even though the mass loss rate prescribed in eq. \ref{equn:evap} can be calculated for any size disc, due to the smoothing function applied by eq. \ref{eq:smooth}, the actual mass loss rate quickly drops to 0 when the outer disc radius is located interior to $\rgfuv$.

\subsection{Simulation Parameters}
The main parameters we vary in this work are the rates of internal and external photoevaporation.
For the external photoevaporative mass loss rates, we vary the initial mass-loss rates for a 100 $\au$ disc, ($\dot{M}_{\rm pe,ext,0}$), between $10^{-10} \msunyr$ (e.g. shielded discs, or those evolving in isolation) to $10^{-6} \msunyr$ (e.g. stars evolving in the high UV environment close to O-type stars).
As well as varying the local environment, we also explore stars of different masses: 0.1, 0.3 and 1$\msun$.

X-ray luminosities are observed to vary by up to two orders of magnitude even for stars of the same mass, due to a combination of measurement uncertainty and genuine intrinsic differences in X-ray activity levels, which are time varying \citep[see figure 1 of][ and the associated discussion]{Flaischlen21}.
We therefore account for this spread by considering X-ray luminosities that span two orders of magnitude, centered around $L_X = 10^{[28.5, 29.5, 30.5]}$ for stars of mass 0.1, 0.3 and 1 $\msun$ respectively motivated by \cite{Flaischlen21}.

We also explore the effect of changing the degree of viscous transport of material through the disc by using $\alpha$ viscosity parameter values from  $10^{-4}-10^{-3}$.
Whilst we utilise viscous prescriptions in this work to account for the mass flow through the disc, we stress that it is only used as a proxy for the material transport and that other processes capable of facilitating this \citep[e.g. magnetically driven disc winds;][]{Kunitomo20} should yield similar results and evolutionary tracks.

We assume that the initial disc radius is a function of stellar mass as
\begin{equation}
    r_{\rm ini} = 200 \au \times \left(\frac{M_*}{\msun}\right)^{0.3}.
\end{equation}
This relation corresponds to radii of between 100--200$\au$ for the stellar masses considered here. These radii are consistent with those seen for observed protoplanetary discs, but as yet there is no consensus on the observed initial protoplanetary disc radii as a function of the stellar mass. The relation here is also similar to that used in other works \citep[e.g.][who use an exponent of 0.45 based on the observed disc masses in Ophiuchus\citep{Andrews10}]{wilhelm22}.

For the initial mass of the disc we follow \citet{Haworth20} where from hydrodynamic simulations they find the maximum disc mass $M_{\rm d, max}$ that a gas disc of radius $r_{\rm ini}$ around a star of mass $M_*$ can be before becoming gravitationally unstable is equal to
\begin{equation}
    \label{eq:max_disc_mass}
    \dfrac{M_{\rm d, max}}{M_*} < 0.17 \left(\dfrac{r_{\rm ini}}{100\au}\right)^{1/2}\left(\dfrac{M_*}{\msun}\right)^{-1/2}.
\end{equation}
We take the mass of the discs in the main bulk of this work to be equal to $0.5\times M_{\rm d, max}$.
Table \ref{tab:parameters} shows the simulation parameters for all stellar masses studied in this work.

\begin{table}
    \centering
    \begin{tabular}{c|ccc}
    \hline
    Parameter & 0.1 $\msun$ & 0.3 $\msun$ & 1 $\msun$ \\
    \hline
        $r_{\rm ini} (\au)$ & 100 & 140 & 200 \\
        $M_{\rm d,max} (\msun)$ & 0.054 & 0.11 & 0.24 \\
        $\log_{10}(L_X)$ & 27.5--29.5 & 28.5--30.5 & 29.5--31.5 \\
        $f_{41}$ & 1 & 3 & 10 \\
        $\alpha$ & \multicolumn{3}{c}{[$10^{-4}, 3\times10^{-4}, 10^{-3}$]} \\
        $\dot{M}_{\rm pe, ext, 0} (\msunyr)$ & \multicolumn{3}{c}{$10^{-10}$--$10^{-6}$} \\
    \hline
    \end{tabular}
    \caption{Simulation Parameters. They are, from top to bottom, the initial disc radius, the maximum stable initial disc mass (we use half of this value in our initial conditions), the X-ray luminosity, the EUV ionising flux in unist of $10^{41}\,$s$^{-1}$, the viscous $\alpha$ parameter and the external photoevaporation rate. }
    \label{tab:parameters}
\end{table}

\begin{figure}
    \centering
    \includegraphics[scale=0.6]{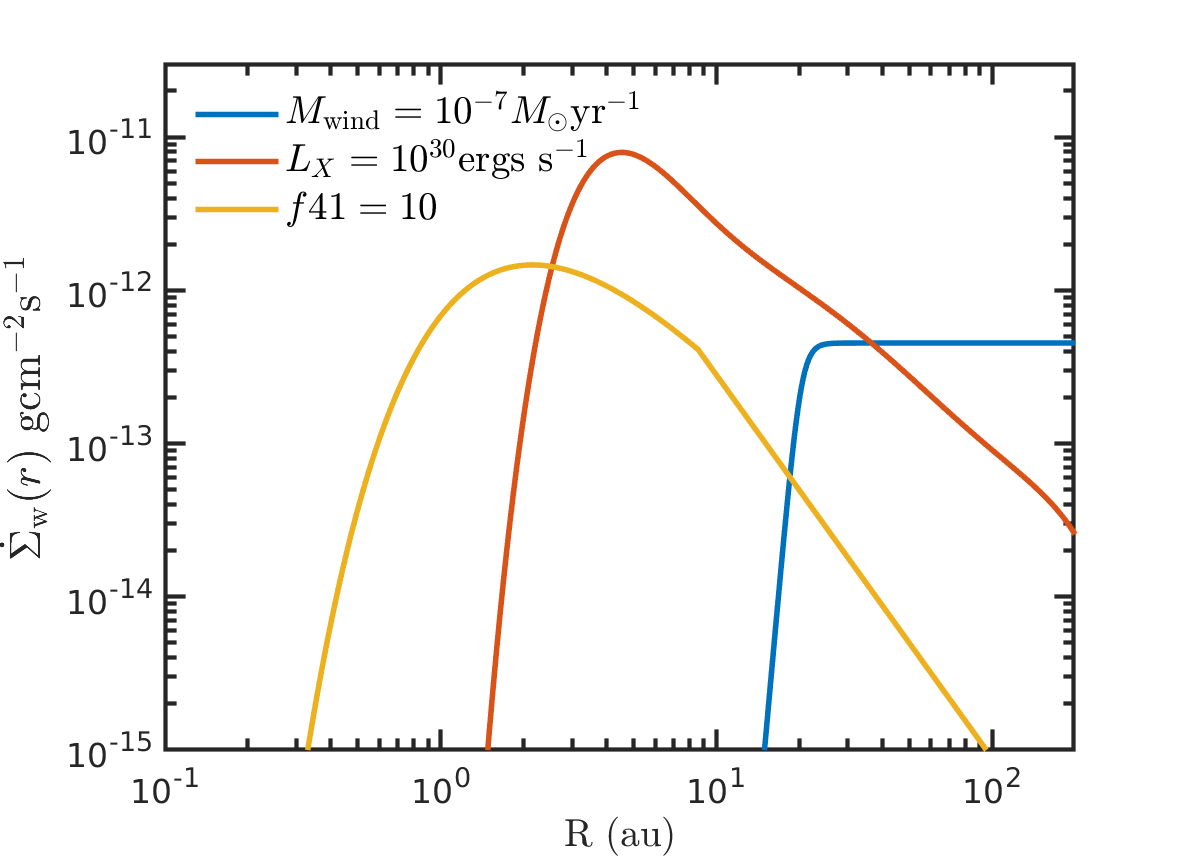}\\
    \caption{Surface density mass-loss rate profile for the different photoevaporation mechanisms. The blue line shows external photoevaporation (eq. \ref{eq:sig_dot_external}) with $\dot{M}_{\rm pe,ext,0}=10^{-7}\msunyr$. The red line shows X-ray dominated internal photoevaporation (eq. \ref{eq:sig_dot_xray}) with $L_X=10^{30}{\rm ergs~s^{-1}}$. The yellow line shows EUV dominated internal photoevaporation (eq. \ref{eq:sig_dot_euv}) with $f_{41}=10$.}
    \label{fig:rates}
\end{figure}

\begin{figure*}
    \centering
    \vspace{-0.5cm}
    \includegraphics[scale=0.42]{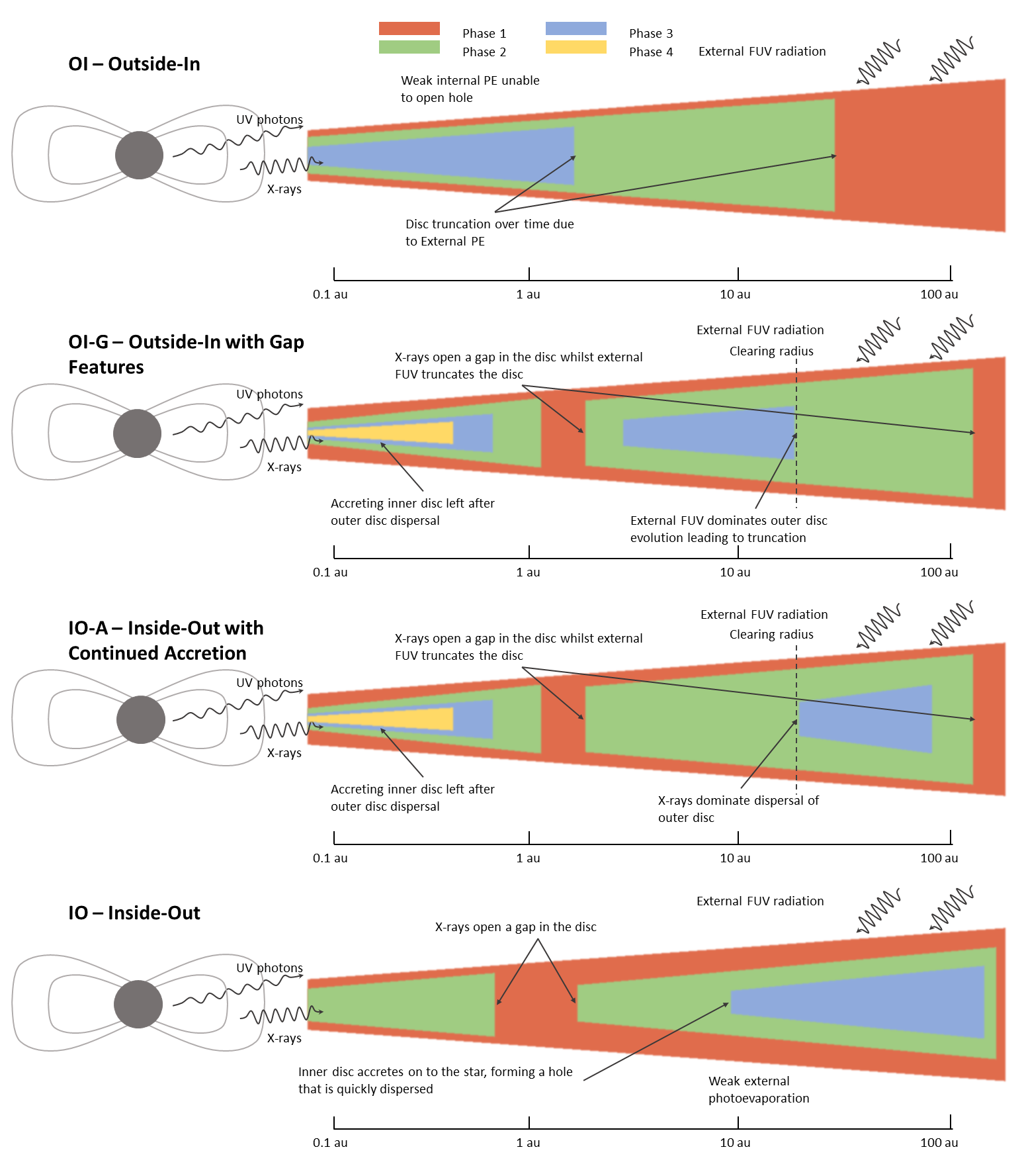}\\
    \caption{Representation of the four main evolution pathways for protoplanetary discs observed from simulations.
    Different coloured regions denote different phases of a specific disc's lifetime.
    A full explanation of the evolution of each disc can be found in sects. \ref{sec:IO}--\ref{sec:OI-G} as well in table. \ref{tab:dispersal_types}.
    We do not include an example for {\it Long-Lived discs} here as they only gradually decrease in size over time, before typically following a pathway similar to {\it Inside-Out (IO)} after 20 Myr.
    The line labelled ``clearing'' radius denotes the point in the disc that we use to distinguish between the IO-A and OI-G scenarios. 
    }
    \label{fig:evolution_cartoon}
\end{figure*}

\subsection{Illustrative comparison of photoevaporation mechanisms }
\label{sec:photoCompare}
Before we examine the evolution of the protoplanetary discs subject to internal/external photoevaporation of varying strengths, it is important to understand the areas of discs where each photoevaporation mechanism dominates.
Figure \ref{fig:rates} shows the  surface density loss rate as a function of radius due to the internal EUV (yellow line), internal X-ray (red line) and external FUV photoevaporation (blue line) for a single combination of parameters.
Internal photoevaporation drives photoevaporative winds from the inner regions of the disc, and quickly falls off farther from the central star.
On the other hand, external photoevaporation is most effective in the outer regions of the disc, and quickly drops off at distances $r<20\au$.
Of course, the relative strengths of these three components will also change somewhat across our parameter space, ultimately determining whether internal or external photoevaporation dominate disc evolution.

\section{Results}
\label{sec:results}

\subsection{Overview of different disc evolution pathways}

The main objective of this work is to examine the different possible final dispersion mechanisms for protoplanetary discs simultaneously undergoing internal and external photoevaporation. We identify 5 different pathways, detailed below, which we refer to as inside-out, outside-in, inside-out with continued accretion, outside-in with gap features and long-lived discs.
In fig. \ref{fig:evolution_cartoon} we illustrate the first 4 pathways mentioned, omitting long-lived discs from the figure as the evolution of these discs, as will be seen in following sections, occurs over a long time-scale before following a pathway qualitatively similar to inside-out evolution.
For each pathway illustrated in fig. \ref{fig:evolution_cartoon}, the colours represent different phases of a disc's lifetime with red being the start of the disc's life and either blue or yellow being the end.
Note that a concise summary of all clearing mechanisms is also given in Table \ref{tab:dispersal_types}.
We will discuss the properties and the temporal evolution of each pathway in the sections below, including the expecting observational parameter space in which they can be found.

\begin{figure}
    \centering
    \includegraphics[scale=0.6]{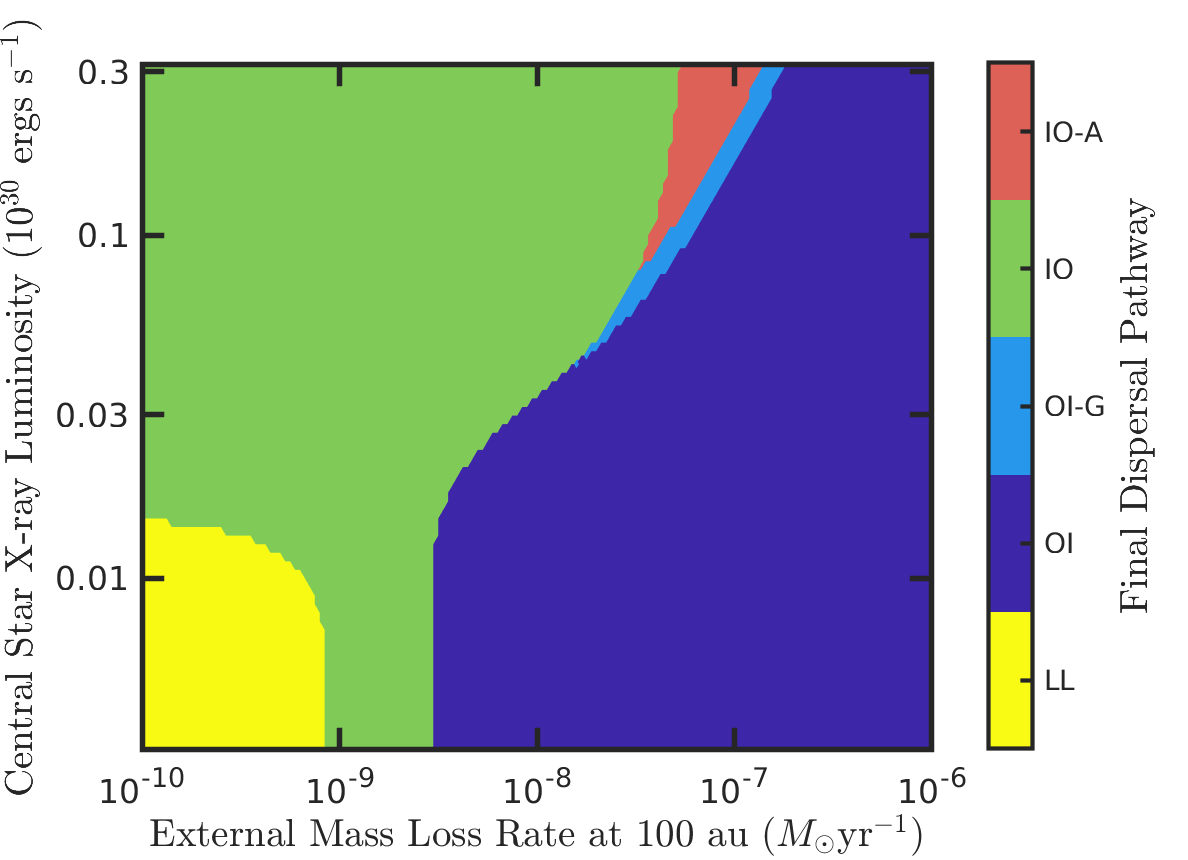}\\
    \caption{Final dispersal pathways for protoplanetary discs around 0.1 $\msun$ mass stars, with a viscosity parameter $\alpha = 3\times10^{-4}$, an initial disc mass equal to half the maximum gravitationally stable disc mass (eq. \ref{eq:max_disc_mass}).
    The final dispersal pathways are as follows: IO-A = {\it Inside-Out with Continued Accretion} , IO = {\it Inside-Out dispersal}, OI-G = {\it Outside-In with Gap features}, OI = {\it Outside-In} and LL = {\it Long-lived discs}.}
    \label{fig:single}
\end{figure}

Fig. \ref{fig:single} shows the outcome for models in our parameter space with 0.1$\msun$ stars,  0.027$\msun$ discs and $\alpha=3\times 10^{-4}$, broken down into the different dispersal pathways. We will now examine each final disc evolution and clearing mechanism in the case of these 0.1$\msun$ stars in more detail, before studying the wider parameter space (i.e. vicosity and stellar mass). 

\subsubsection{Inside-out (IO) dispersal}
\label{sec:IO}
Inside-out disc dispersal probably represents the best studied disc evolution and dispersal paradigm \citep[e.g.][]{2016PASA...33....5O, 2017RSOS....470114E}. Discs evolve through viscous accretion and photoevaporation until the latter stages of their lifetime, when photoevaporation opens an inner gap. Once the innermost region of the disc accretes onto the central star, and the inner disc becomes optically thin, ionising photons launch a wind directly off the inner edge of the disc, quickly evaporating it from the {\it inside-out} (\textit{IO}).

\begin{figure}
    \centering
    \includegraphics[scale=0.6]{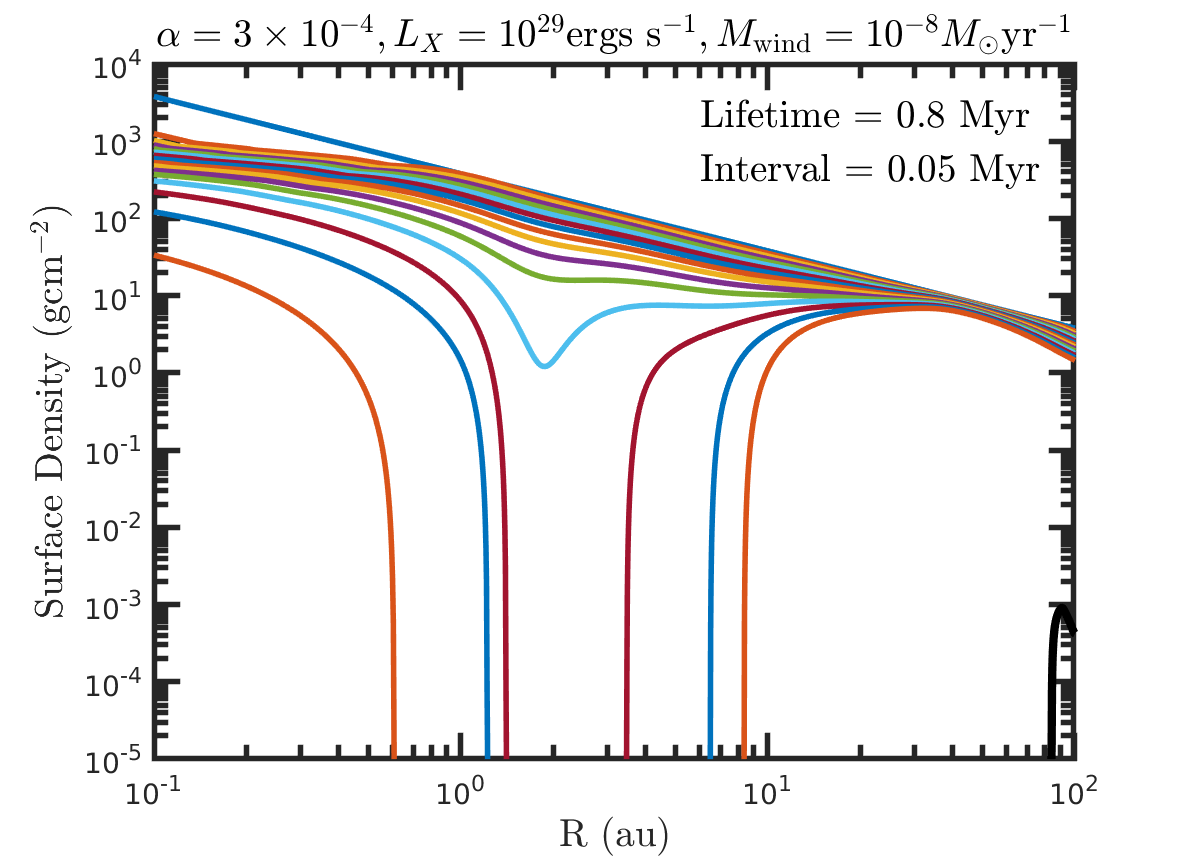}\\
    \caption{Surface density profiles of an evolving protoplanetary disc around a 0.1$\msun$ star undergoing {\it Inside-Out (IO)} dispersal at the end of the disc lifetime. The uppermost blue line shows the initial disc profile, whilst the black line shows the last disc output before the disc had fully dispersed. The surface density is shown in intervals of intervals of 0.05 Myr.}
    \label{fig:discs_IO}
\end{figure}

The surface density evolution of an example disc undergoing IO dispersal is given in fig. \ref{fig:discs_IO}. 
The disc initially had a mass equal to half the maximum gravitationally stable mass (eq. \ref{eq:max_disc_mass}), with a viscous $\alpha=3\times 10^{-4}$, a central stellar luminosity of $L_X=10^{29} \rm ergs~s^{-1}$, and was situated in an environment where the mass loss rate at $100\au$ was $M_{\rm pe,ext,0}=10^{-8}\msunyr$.
The profiles in the disc evolve from the top of the figure to the bottom, as the disc is continually losing mass.
Internal photoevaporation opens a gap in the disc at $\sim2\au$, after $\sim0.6$ Myr, leaving the outer disc to continue to lose mass through photoevaporation, and the inner disc to accrete onto the central star.
Due to the weak external photoevaporation rate, a substantial amount of gas is still present far from the star, as the gap opened.
This abundance of material is sufficient to allow the inner disc time to accrete onto the star, before the outer disc is lost through photoevaporation.
After 0.79 Myr, the majority of the inner disc is accreted onto the central star, leaving the inner disc optically thin.
The central star then drives a photoevaporative wind off of the inner edge of the outer disc, dramatically increasing the magnitude of photoevaporation, quickly removing the outer disc from the {\it inside-out}.

In fig. \ref{fig:single}, the green area denotes where {\it IO} operates, i.e. where the external photoevaporation rates are low, allowing the gas in the outer disc to survive longer than the inner disc where internal photoevaporation operates.
The green area ({\it IO}) also occurs for stars with low X-ray luminosities and moderate values of external photoevaporation due to EUV-driven internal photoevaporation. Moderate external photoevaporative mass loss rates limit the supply of material to the inner disc, meaning EUV photons are able to open gaps in the inner disc, before fully opening holes and triggering inside-out evolution.

\subsubsection{Outside-in (OI) dispersal}
\label{sec:OI}
When external photoevaporation dominates the disc evolution, it rapidly truncates the disc all the way down to the effective gravitational radius ($\beta\rgfuv$).
This truncation of the outer disc stems the flow of gas into the region where internal photoevaporation operates, thus reducing the impact of internal photoevaporation on the disc, as well as the amount of time it has to open a gap in the disc.
Even though external photoevaporation does not operate in the inner regions of the disc, it still dominates the evolution of the disc, as the remaining disc mass continues to accrete on to the central star, with the outermost gas viscously spreading, and being photoevaporated by the external radiation.
Rather than opening a gap in the disc, the gas surface density gradually drops as the disc loses mass, reducing in size over time.
The disc is fully dispersed once the inner disc has accreted on to the star finalising the outside-in evolution pathway.

The dark blue region in fig. \ref{fig:single} shows the regions where {\it outside-in (OI)} dispersal dominates, which is typically for an external mass loss rate normalisation of $\geq10^{-7}\,$M$_\odot$\,yr$^{-1}$ (corresponding to about $10^4$\,G$_0$), though can go as low as around $\sim$3$\times10^{-9}$M$_\odot$\,yr$^{-1}$  (corresponding to about $300$\,G$_0$). 
Intuitively, these regions would correspond to high FUV radiation field environments such as in the vicinity of massive stars in regions like the central Orion Nebular Cluster (ONC). 
As the X-ray luminosity of the central star increases, the region where external radiation dominates shrinks, since there is significant competition for material in the intermediate region of the disc, which allows a gap to open in the disc, thus a different evolution pathway to the disc consistently truncating down.
This alternative pathway will be discussed in sect. \ref{sec:OI-G}.

\begin{figure}
    \centering
    \includegraphics[scale=0.6]{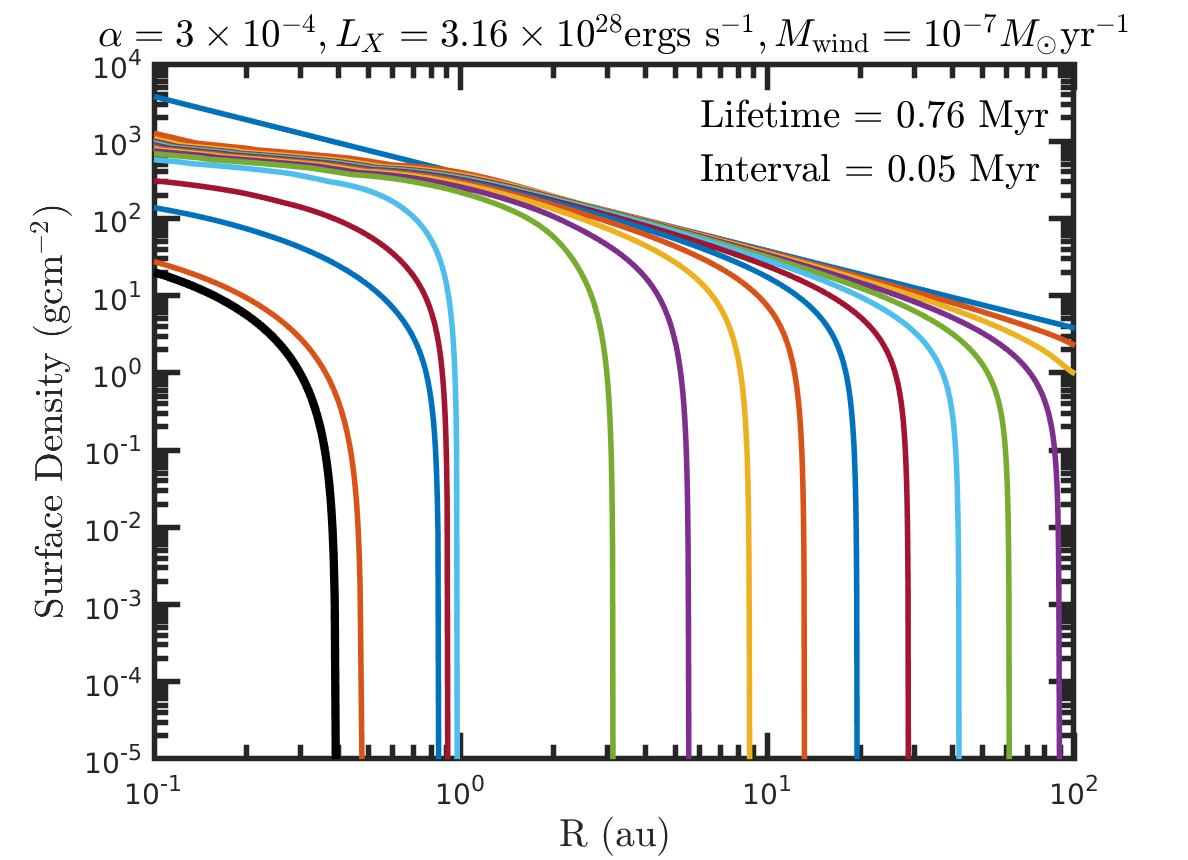}\\
    \caption{Surface density profiles of an evolving protoplanetary disc around a 0.1$\msun$ star undergoing {\it Outside-In (OI)} dispersal at the end of the disc lifetime. The uppermost blue line shows the initial disc profile, whilst the black line shows the last disc output before the disc had fully dispersed. The surface density is shown in intervals of intervals of 0.05 Myr.}
    \label{fig:discs_OI}
\end{figure}

An example of the surface density evolution for outside-in dispersal is given in fig. \ref{fig:discs_OI}.
Like the example for inside-out evolution (fig. \ref{fig:discs_IO}), this example has an initial mass equal to half the gravitationally stable mass (eq. \ref{eq:max_disc_mass}), a viscous $\alpha=3\times10^{-4}$, a central stellar X-ray luminosity equal to $3\times10^{28}$ ergs s$^{-1}$, with an initial external mass loss rate normalisation of $10^{-6}\msunyr$.
For this disc, the gas surface density depletes and truncates over time from the outside-in.
Note, that the X-ray luminosity of the central star in this example was equal to that of the disc shown in fig. \ref{fig:discs_IO}, but internal clearing did not dominate in this case because the external clearing was so much stronger.
The truncation of the disc continued to around ~$\sim 2\au$, the region where external photoevaporation is ineffective around such low mass stars.
This left the inner disc to accrete on to the central star, further truncating the disc as it evolved and came to the end of its lifetime.

\subsubsection{Inside-out dispersal with continued accretion (IO-A)}
\label{sec:IO-A}

The \textit{IO} and \textit{OI} mechanisms described above represent the extreme scenarios of the internal or external mass loss mechanisms dominating the dispersal of protoplanetary discs.
However there are also intermediate dispersal regimes incorporating elements of both pathways.
One such is \textit{inside-out dispersal with continued accretion} (\textit{IO-A}) where the main mode of disc evolution is similar to inside-out dispersal, with internal photoevaporation being the dominant source of photoevaporative mass loss.
In this scenario, internal photoevaporation opens a gap and pushes the inner edge of the outer disc further from the central star.
If the accretion rate through the disc, i.e. viscosity, is weak enough, then the time-scale for the inner disc to accrete on to the central star will be longer than the clearing time-scale of the outer disc through photoevaporation.
With internal photoevaporation being dominant over external photoevaporation, the outer disc will continue to retreat outwards, with the gap becoming larger in size, similar to the late stages of inside-out dispersal.
This continues until the outer disc has been fully dispersed, but the inner disc remains, accreting on to the central star, making the disc appear as if it has been truncated down by external photoevaporation, but instead it has been dispersed through strong internal photoevaporation.

To differentiate whether the outer disc has mainly been dispersed by internal or external photoevaporation, we define an arbitrary ``clearing radius'' that is equal to $5\times\beta\rgfuv$, with the value determined empirically by inspection of the behaviour of our models across the parameter space.
Note that the use of this radius is only to allow the determination of which mechanism dominates to be as consistent as possible across all stellar masses, and since one of internal or external clearing usually strongly dominates, using a modestly different clearing radius value does not affect the results substantially. 

For discs undergoing the {\it IO-A} evolution pathway, internal photoevaporation is able to clear the outer disc interior to the clearing radius, before external photoevaporation can truncate the remaining outer disc down to the clearing radius (see the middle panels of fig. \ref{fig:evolution_cartoon} for an illustration of the clearing radius and these evolution pathways).

In fig. \ref{fig:single}, the red region shows the parameters where the disc evolution follows the inside-out with continued accretion pathway, denoted as {\it IO-A}.
For the simulated choice of discs and viscosity, this region appears in between the inside-out region (green), and regions favouring outside-in and outside-in with gap features (light and dark blue).
In this region, external photoevaporation has been able to slightly truncate the disc, but not significantly.
This is what allows internal photoevaporation to disperse the outer disc before the inner disc can accrete on to the star.
To the region left of this (green, {\it IO}), external photoevaporation was weaker and thus had not been able to remove sufficient amounts of gas, leaving the gas there for internal photoevaporation to disperse.
This gave the inner disc enough time to accrete on to the central star and create a large hole in the disc, resulting in the traditional form of {\it inside-out} dispersal.
We will discuss the region to the right of the {\it IO-A} region in the section below.

\begin{figure}
    \centering
    \includegraphics[scale=0.6]{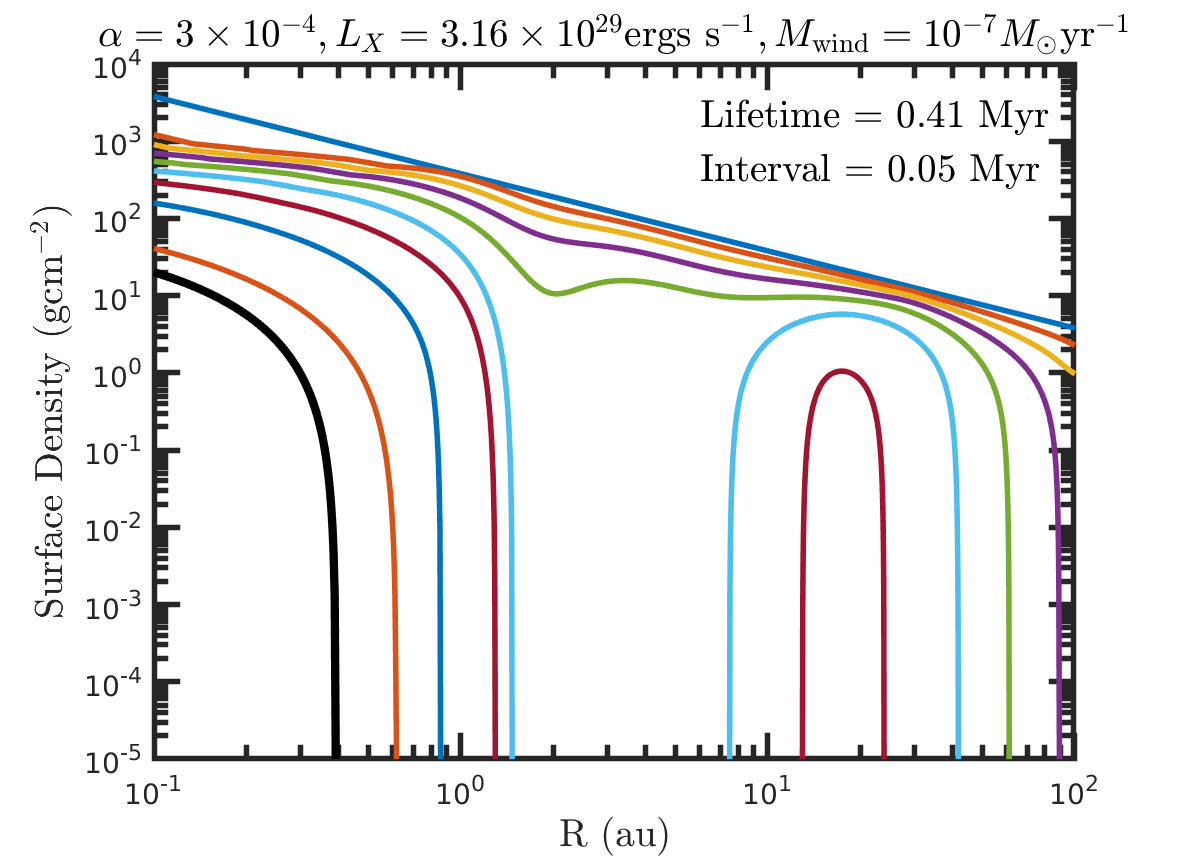}\\
    \caption{Surface density profiles of an evolving protoplanetary disc around a 0.1$\msun$ star undergoing {\it Inside-Out with Continued Accretion (IO-A)} evolution throughout the disc lifetime. The uppermost blue line shows the initial disc profile, whilst the black line shows the last disc output before the disc had fully dispersed. The surface density is shown in intervals of intervals of 0.05 Myr.}
    \label{fig:discs_IO-A}
\end{figure}

Figure \ref{fig:discs_IO-A} shows the surface density profiles for a disc experiencing the {\it IO-A} evolution, where the disc had a lifetime of 0.41 Myr.
The surface density profiles are shown for every 0.05 Myr, with the black line showing the final profile.
For this disc, the X-ray luminosity of the central star was equal to $3.16\times10^{29}$ ergs s$^{-1}$, and the external photoevaporation rate was equal to $10^{-7}\msunyr$.
As can be seen by the green and cyan profiles in the middle of fig. \ref{fig:discs_IO-A}, internal photoevaporation has been able to open a gap in the disc after $\sim0.21$ Myr, whilst external photoevaporation has been able to truncate the disc down to $\sim55~\au$.
During the remaining disc lifetime, the outer disc lost mass through both internal and external photoevaporation, with the inner disc accreting on to the central star.
For this disc, internal photoevaporation was able to open a significant gap in the disc before external photoevaporation could truncate it.
After $\sim 0.35$ Myr, the outer disc had been completely dispersed leaving the inner disc to accrete on to the central star whilst also undergoing slow EUV photoevaporation from the central star, with the disc eventually fully dispersed after 0.43 Myr.

\subsubsection{Outside-in dispersal with gap features (OI-G)}
\label{sec:OI-G}

As discussed in \ref{sec:IO-A}, there is a region in the parameter space between {\it IO} and {\it OI} dispersal where internal photoevaporation is able to open a gap in the disc.
We now look at the analogous intermediate regime where external photoevaporation dominates.
In this scenario external photoevaporation acts to moderately truncate the disc, but internal photoevaporation is also strong enough to cause significant mass loss and open a gap in the disc before external photoevaporation can fully truncate the disc.
With a gap formed, the outer disc is then cleared from both sides by both internal and external photoevaporation. External photoevaporation truncates down to the clearing radius first in this regime, being the main driver of dispersal in the outer disc, this leads us to denote the mechanism as {\it outside-in with gap features}, since sizeable gaps are able to form before the disc is fully truncated.

\begin{figure}
    \centering
    \includegraphics[scale=0.6]{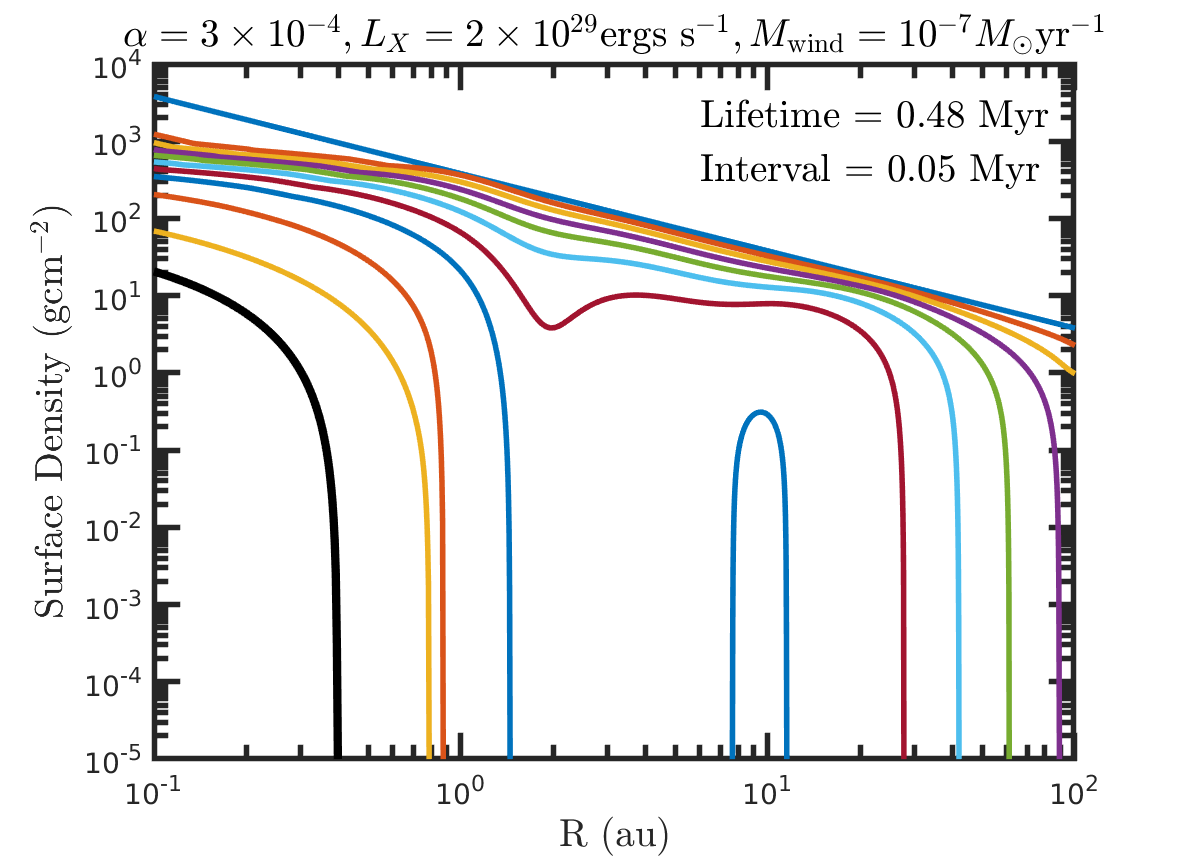}\\
    \caption{Surface density profiles of an evolving protoplanetary disc around a 0.1$\msun$ star undergoing {\it Outside-In with Gap features (OI-G)} evolution. The uppermost blue line shows the initial disc profile, whilst the black line shows the last disc output before the disc had fully dispersed. The surface density is shown in intervals of intervals of 0.05 Myr.}
    \label{fig:discs_OI_G}
\end{figure}

Figure \ref{fig:discs_OI_G} shows the surface density profiles of a disc around a low-mass star where the evolution of the disc follows the {\it OI-G} pathway. 
The X-ray luminosity ($L_X=2\times10^{29}$ergs s$^{-1}$) is similar to the examples for {\it IO} example (fig. \ref{fig:discs_IO}) and {\it OI} example (fig. \ref{fig:discs_OI}), but with an external photoevaporation rate between the two ($\dot{M}_{\rm pe,ext,0}=10^{-7}\msunyr$), hence leading to an intermediate evolution pathway incorporating the effects from both internal and external photoevaporation.
With the X-ray luminosity being slightly reduced compared to the {\it IO-A} example (fig. \ref{fig:discs_IO-A}), external photoevaporation is able to disperse a greater amount of gas in the outer disc, truncating it significantly as a gap opens after $\sim0.3$ Myr, before the outer disc fully disperses after $\sim0.4$ Myr, leaving the inner disc to accrete on to the central star.

The light blue region in fig. \ref{fig:single} shows the region where outside-in with gap features ({\it OI-G}) is the mode of disc evolution.
This region is indeed situated between the {\it IO} (green) and {\it OI} (dark blue) evolution pathways, with central stars exhibiting significant X-ray luminosities, whilst also being subjected to moderate external radiation fields.

\subsubsection{Long-lived (LL) discs}
\label{sec:LL}

The final evolution pathway arises when the mass loss rates from internal and external photoevaporation are small, and comparable to the mass flow through the disc.
This results in limited amounts of gas being lost through photoevaporation, which allows the disc to slowly accrete on to the central star.
For our simulations, we allow them to evolve for 20 Myr, and if there is still gas left at the end of the simulations ($>0.1\me$), then they are considered to be long lived.
In some scenarios, the discs will be near the end of their lifetime, where their evolution may be about to follow an inside-out type pathway.
However for other discs, where accretion on to the central star is slow, and photoevaporation rates are weak, these discs will still be containing significant amounts of gas and may possess much longer disc lifetimes than those studied here.

In fig. \ref{fig:single}, the long-lived discs, denoted {\it LL}, are shown in yellow, and unsurprisingly exist in the regime where the X-ray luminosity from the central star and external radiation from nearby sources are weakest. 

\begin{figure}
    \centering
    \includegraphics[scale=0.6]{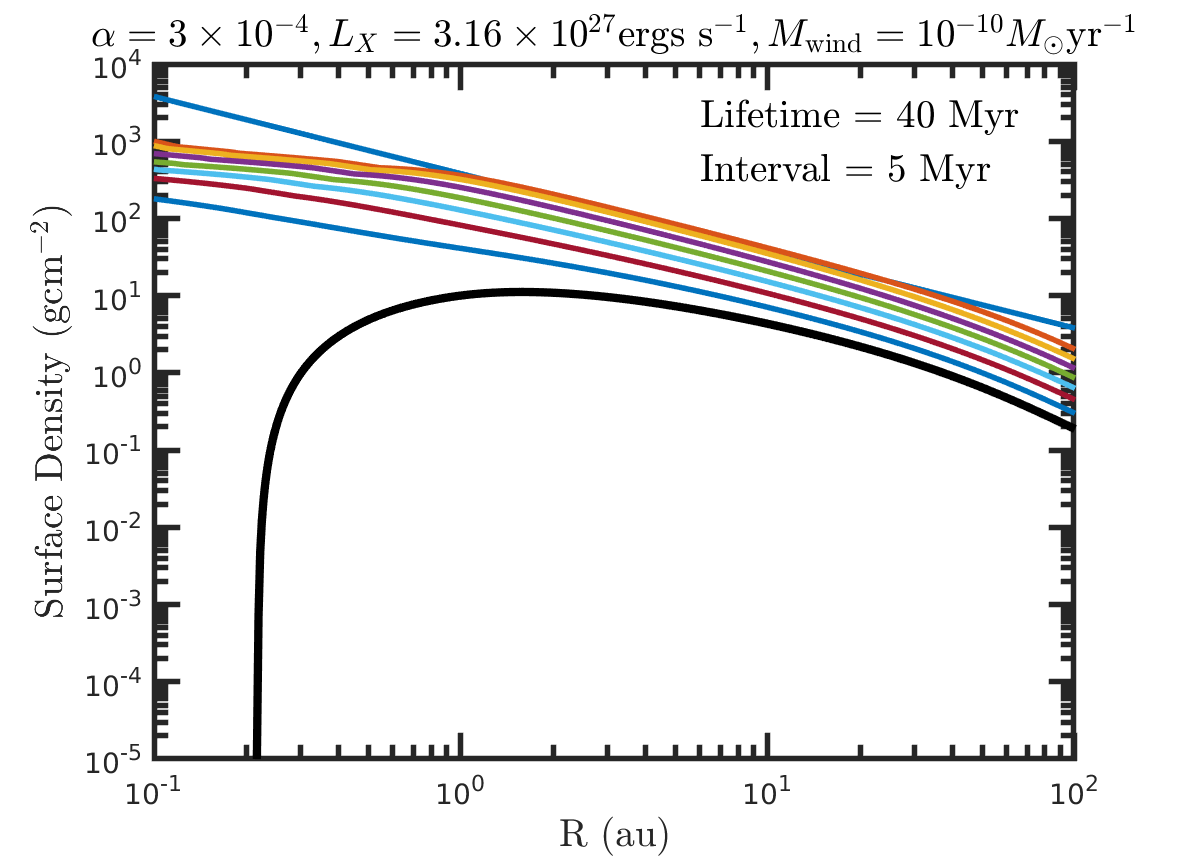}\\
    \caption{Surface density profiles of an evolving {\it Long-Lived }protoplanetary disc around a 0.1$\msun$ star, with a lifetime of 40 Myr. The uppermost blue line shows the initial disc profile, whilst the black line shows the last disc output before the disc has fully dispersed. The surface density is shown in intervals of intervals of 5 Myr.}
    \label{fig:discs_LL}
\end{figure}

We show the surface density profiles for an example long-lived disc in fig. \ref{fig:discs_LL} where the disc ultimately had a lifetime equal to 40\,Myr (we allowed this one to evolve for longer than 20\,Myr).
As can be seen by the surface density profiles, the mass lost due to X-rays from the central star as well as external radiation had little impact over the disc lifetime.
Only near the end of the disc lifetime, when the gas surface density was significantly reduced, was internal EUV photoevaporation able to open a hole in the inner regions of the disc, and trigger the final stage of it's evolution - dispersing from the inside-out.
With the disc lifetime, being longer than 20 Myr, this is the reason the disc is classified as {\it long-lived}, as opposed to {\it IO}.

\subsection{Disc Lifetimes}
\label{sec:lifetimes}
Whilst knowing the mode of disc evolution is important, it is also interesting to examine the lifetimes of discs as the properties of their host star and local environment vary.
Through observing disc fractions around local clusters of different ages, typical protoplanetary discs are expected to have lifetimes ranging from 1--10 Myr \citep{Haisch01,Fedele10,Richert18}, with exponential mass accretion time-scales of between 2--3 Myr.
There is also evidence that protoplanetary discs around lower mass stars have longer lifetimes than those around higher mass stars \citep{Ribas15}, but this seems to only be observable for clusters with ages greater than 1 Myr.

\begin{figure}
    \centering
    \includegraphics[scale=0.6]{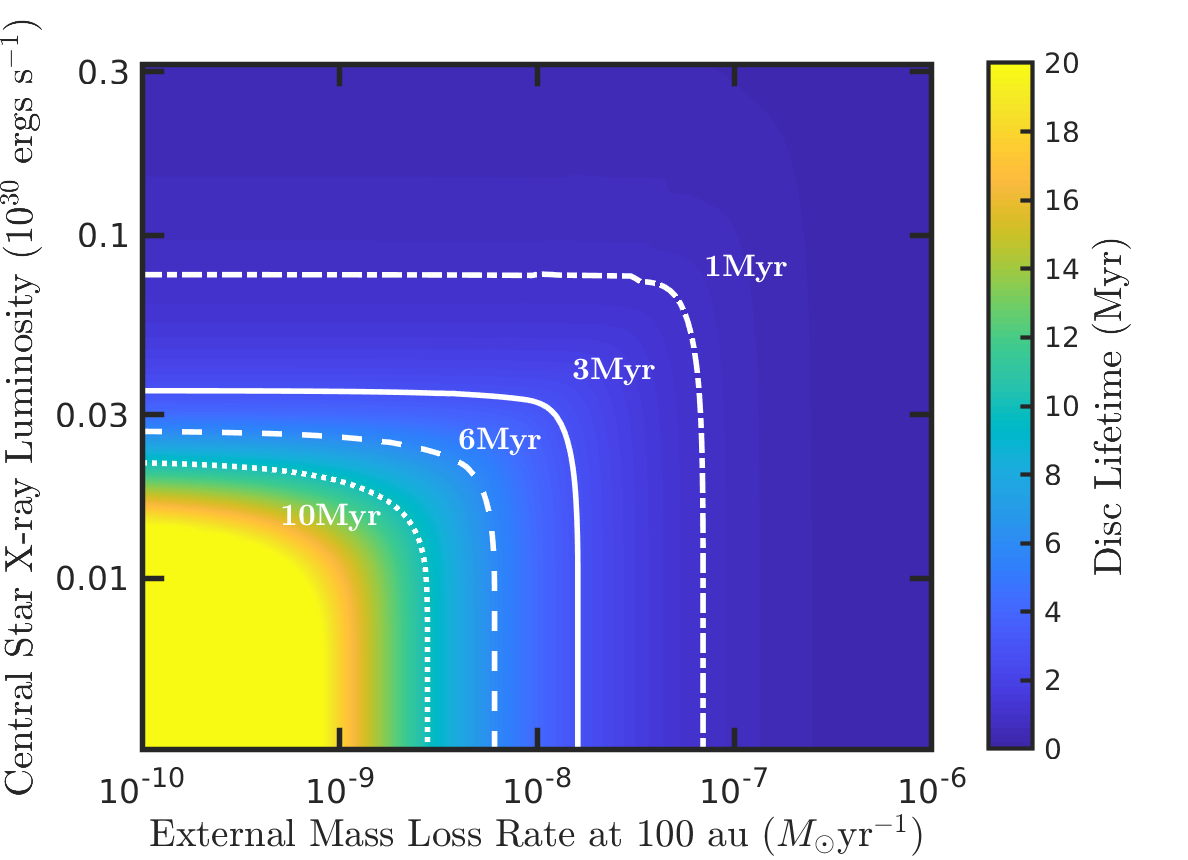}\\
    \caption{Contour plot showing the lifetimes of discs around a 0.1 $\msun$ star with a viscosity parameter $\alpha=3\times 10^{-4}$, and an initial disc mass equal to half the maximum stable disc mass (eq \ref{eq:max_disc_mass}). The white lines show contours equal to 1 Myr (dot-dashed), 3 Myr (solid), 6 Myr (dashed) and 10 Myr (dotted).}
    \label{fig:lifetime}
\end{figure}

Figure \ref{fig:lifetime} shows the lifetimes of the discs that were presented in fig. \ref{fig:single}, orbiting a 0.1$\msun$ star with the viscosity parameter $\alpha=3\times 10^{-4}$.
The white lines show the contours for disc lifetimes of 1 Myr (dot-dashed), 3 Myr (solid), 6 Myr (dashed) and 10 Myr (dotted).
Unsurprisingly, the disc lifetimes are longest when the X-ray luminosity and the external mass loss rate are weakest, as shown by the yellow contours in the bottom-left of fig. \ref{fig:lifetime}. Interestingly when either photoevaporation regimes are at their strongest, disc lifetimes are found to be extremely short, less than 1 Myr, and for the strongest parameters studied here, less then 0.5 Myr which could prohibit even very early planet formation \citep[e.g.][]{2020Natur.586..228S}.
From observations of disc fractions in local clusters typical protoplanetary disc lifetimes are expected to be longer around low-mass stars than higher mass stars \citep{Ribas15}, but such observations also suggest that the fractions are independent of stellar mass for the first 1 Myr.
Therefore, whilst the short-lived discs found here may have lifetimes much shorter than expected around low-mass stars, they may still exist in the observations where the disc fractions are independent of stellar mass.

\subsection{Comparison of mass lost through different photoevaporation mechanisms}
\label{sec:ratios}

\begin{figure}
    \centering
    \includegraphics[scale=0.6]{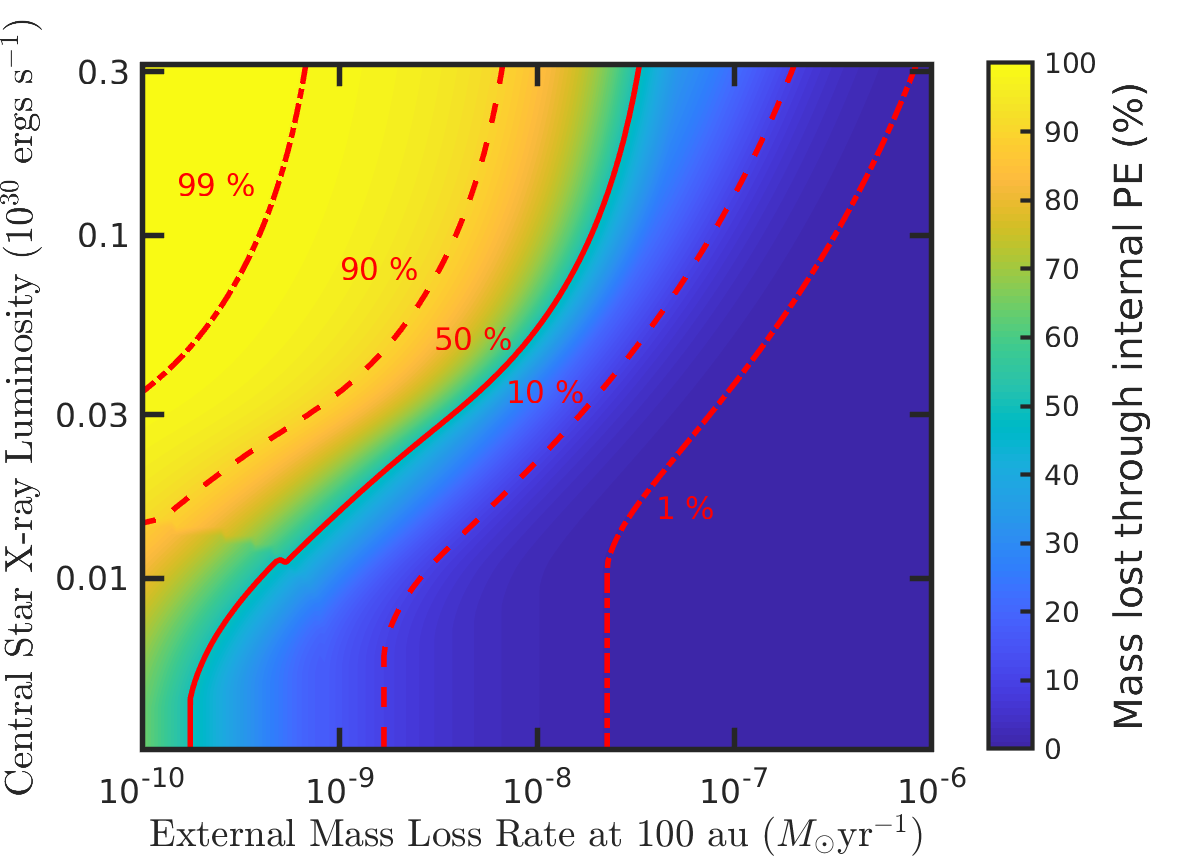}\\
    \caption{Contour plot showing the percentage of mass lost through internal photoevaporation (ignoring accretion) around a 0.1 $\msun$ star with a viscosity parameter $\alpha=3\times 10^{-4}$, and an initial disc mass equal to half the maximum stable disc mass (eq \ref{eq:max_disc_mass}). Red lines indicate from right-to-left: 1\% (dot-dashed), 10\% (dashed), 50\% (solid), 90\% (dashed) and 99\% (dot-dashed).}
    \label{fig:evap_ratio}
\end{figure}

With this work focusing on the evolution of protoplanetary discs undergoing both internal and external photoevaporation, it is interesting to compare the total mass lost from both mechanisms and show where each mechanism is dominant and how they connect to the disc evolution pathways.
In fig. \ref{fig:evap_ratio} we show the percentage of the mass lost in the disc through internal photoevaporation (excluding accretion).
The red lines in fig. \ref{fig:evap_ratio} show contours where the percentages of mass lost through internal photoevaporation are equal to: 1\%, 10\%, 50\%, 90\% and 99\%.
As expected, the region of the parameter space where internal photoevaporation is most dominant is the top-left corner of fig. \ref{fig:evap_ratio} where the X-ray luminosity is strongest, and the external mass loss rate is at its lowest.
This region is fully embedded in the inside-out evolution pathway with this pathway also being able to extend out past the 50\% line where the mass lost through internal and external photoevaporation was equal.

The dark blue region in fig. \ref{fig:evap_ratio} denotes where external photoevaporation dominates internal photoevaporation, resulting in the outside-in evolution pathway.

\section{Varying Stellar mass and Viscosity}
\label{sec:main_plots}

So far we have only examined discs around 0.1$\msun$ stars and with a viscous parameter $\alpha=3\times10^{-4}$, determining the pathways of evolution that discs followed when subjected to varying background UV fields, and X-ray luminosities from their parent stars.
We will now examine how the propensity of these pathways within the parameter space are affected as the stellar mass, and $\alpha$ are varied.

\begin{figure*}
    \centering
    \vspace{-0.3cm}   
    \includegraphics[width=\textwidth]{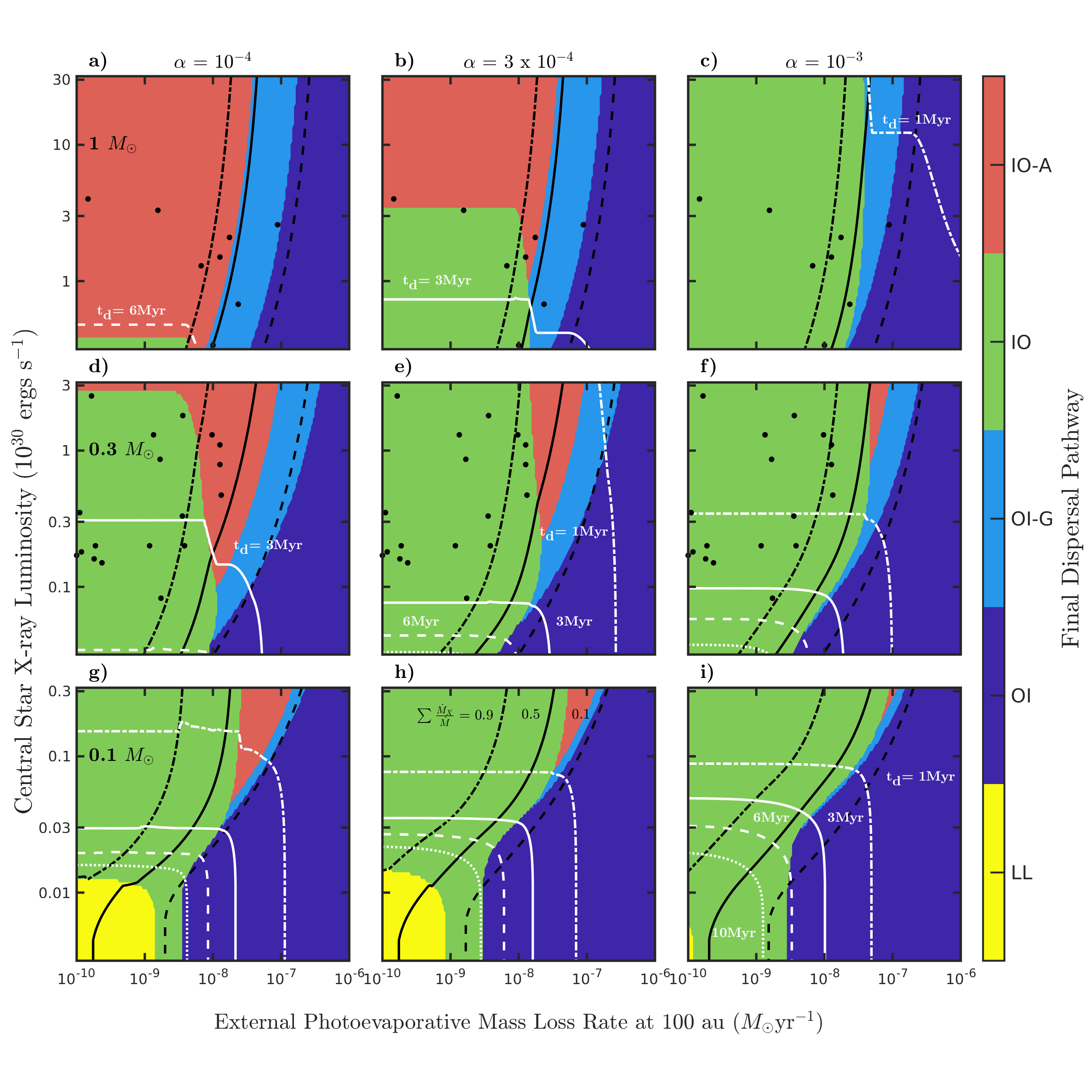}
    \vspace{-1cm}
    \caption{Final dispersal pathways for protoplanetary discs around stars of different masses: 1 $\msun$ (top panels), 0.3 $\msun$ (middle panels), and 0.1 $\msun$ (bottom panels).
    Initial disc masses for these discs was set to half of the maximum stable mass (eq. \ref{eq:max_disc_mass}).
    The effects of changing the viscosity parameter $\alpha$ is also shown with: $\alpha = 10^{-4}$ (left-hand panels), $\alpha = 3\times 10^{-4}$ (middle panels), and $\alpha = 10^{-3}$ (right-hand panels),
    The final dispersal types are as follows: IO-A = {\it Inside-Out with accreting inner disc} , IO = {\it Inside-Out dispersal}, OI-G = {\it Outside-In with Gaps}, OI = {\it Outside-In} and LL = {\it Long-lived discs}.
    Black lines show the fraction of the mass lost through internal photoevaporation compared to the total mass mass lost: dot-dashed = 0.9, solid = 0.5, dashed = 0.1.
    Lifetimes of the discs are shown by the white lines: dot-dashed = 1 Myr, solid = 3 Myr, dashed = 6 Myr, dotted = 10 Myr.
    We denote the locations of observed transition discs with black points using data from \citet{Kim13}.}
    \label{fig:multiplot}
\end{figure*}

Figure \ref{fig:multiplot} presents a group of contour plots showing different evolution pathways and the regions of parameter space they occupy when varying the central stars X-ray luminosity (y-axis) and the radiation emanating from external stars in the local environment (x-axis).
The rows correspond to host star masses of, 0.1, 0.3 and 1$\msun$ from top to bottom. 
The columns correspond to viscous  $\alpha$ parameters of $10^{-4}, 3\times 10^{-4}$ and $\alpha=10^{-3}$ from left to right.
Note the bottom centre panel is identical to fig. \ref{fig:single}.

In each panel, we also overlay contours showing the disc lifetimes (white lines), as well as the ratio of mass lost through internal photoevaporation compared to that lost through all photoevaporation mechanisms (black lines).
For the disc lifetimes we add contours for 1 Myr (dot-dashed), 3 Myr (solid), 6 Myr (dashed), and 10 Myr (dotted).
For the ratio of mass lost through internal photoevaporation we add contours for ratios of 0.9 (dot-dashed), 0.5 (solid), and 0.1 (dashed).

The black points on each plot correspond to the parameters of observed transition discs in the Orion A star forming region \citep{Kim13}, assuming that a disc in an environment with a $G_0$ field of $10^{4}$ is equivalent to an external mass loss rate of a 100 $\au$ disc equal to $3\times10^{-7}\msunyr$.
We again highlight that the transition discs placed here show signatures of gaps with their SEDs, and we make no distinction between observed discs with full inner cavities, and those with gap signatures whilst still accreting on to their central stars.
For each model disc, the initial mass was set to half of the maximum stable disc mass (eq. \ref{eq:max_disc_mass}), this being equal to 0.027$\msun$, 0.055$\msun$, and 0.12$\msun$ for 0.1$\msun$, 0.3$\msun$, and 1$\msun$ stars respectively.

Note that while fig. \ref{fig:multiplot} provides a useful overview of where different dispersal mechanisms dominate, we caution against considering a given classification in absolute terms. For example, just because a disc is classed as clearing by inside-out doesn't mean that negligible mass was lost due to external photoevaporation, and vice versa. 

Next in \ref{sec:low_mass_stars} we examine the effect that different $\alpha$ values have on the final disc dispersal pathways in the 0.1\,M$_\odot$ star case.
Then in \ref{sec:stellar_mass} we analyse the dependence on stellar mass, and determine whether observed transition discs can be explained with current photoevaporation models based on their observed properties.

\subsection{Changing viscosity around low-mass stars}
\label{sec:low_mass_stars}

Here we compare panels g), h) and i) from the bottom row of fig. \ref{fig:multiplot}, which show the final dispersal pathways of discs around 0.1$\msun$ stars with viscous $\alpha$ parameters equal to $10^{-4}$, $3\times 10^{-4}$ and $10^{-3}$ respectively.
The larger $\alpha$ values effectively suppress the existence of long-lived ($>20$\,Myr) discs.
This is due to considerably larger quantities of gas being removed from the disc through accretion on to the central star.

Higher viscosity mainly aids internal photoevaporation as it allows gaps to form in the disc around the effective gravitational radius at an earlier time, before the inner disc quickly accretes on to the central star.
With the hole then forming at an earlier time for the larger $\alpha$ discs, these are then able to be classified as following the inside-out pathway as opposed to being long-lived.
This is especially clear in panel i) showing the discs with $\alpha=10^{-3}$, where only the discs in the weakest of internal and external environments are able to be long-lived.

The other main feature that differs as $\alpha$ is increased is the tendency for discs with high stellar X-ray luminosities to follow the {\it IO} evolution pathway as opposed to {\it IO-A}.
This is not unexpected since with higher values of $\alpha$, the inner disc is able to accrete on to the star much more quickly, allowing holes to form before the outer discs have been lost through photoevaporation.

With increasing $\alpha$ viscosity, the fraction of gas lost through internal photoevaporation is also slightly increased.
This is illustrated by the black lines shifting slightly to the right moving from panels g) to i).
The increase in the ratio is due to the increased gas flow through the disc, resulting from the increase in viscosity.
This reduces the amount of gas available to be lost through external photoevaporation, whilst simultaneously increasing the supply available for internal photoevaporation.
The transfer of gas also has the effect of slightly increasing the frequency of discs following an {\it OI} pathway, since the outer disc is able to be truncated at a slightly faster rate, whilst the area of the disc undergoing internal photoevaporation also contains a larger supply of gas that it needs to remove, further allowing the {\it OI} pathway to be more prominent.

\subsection{Dependence on stellar mass}
\label{sec:stellar_mass}

Here we explore the dependency of dispersal pathway on the host star mass. 
Looking again at fig. \ref{fig:multiplot}, we now examine the middle and top rows showing the evolution pathways of discs around $0.3\msun$ and $1\msun$ stars respectively.

Of course there is a similar dependency of the disc lifetime on the degree of internal and external photoevaporation. However, an immediate, obvious difference is the lack of long-lived discs around the higher mass stars compared to $0.1\msun$ stars, across all values of $\alpha$ studied here. This is due to the increase in X-ray photoevaporation rates in the inner regions of the discs. This is illustrated by the locations of the dot-dashed and solid black lines in panels d) and e) with those in g) and h).
The higher X-ray driven photoevaporation rates at higher stellar masses gaps when the gas surface densities have been sufficiently diminished, i.e. after 10 Myr of evolution as shown by the dotted white lines in panels d) and e).
For solar-mass stars, this effect is even more pronounced with all discs having lifetimes less than 10 Myr, as internal photoevaporation dominates their evolution and is quickly able to open holes in the discs in weak external environments.

In regards to disc lifetimes, it is also clear that in general as the stellar mass increases, mean disc lifetimes decrease.
The mean disc lifetime for 0.1$\msun$ stars in discs with a viscous $\alpha=10^{-3}$ is 3.3 Myr, whilst it is equal to 1.75 Myr and 1.43 Myr for 0.3 $\msun$ and 1 $\msun$ stars respectively.
The reduction in disc lifetimes is due to the increase in mass lost through X-ray photoevaporation, as well as external photoevaporation due to the more massive and extended discs simulated around more massive stars.
Note that the values for the mean disc lifetimes are calculated to show the general trend in lifetime with stellar mass, and not as a derived value from an initial population distribution. This general behaviour is consistent with observations of disc fractions for stars of different mass \citep[e.g.][]{Ribas15}. 

However, whilst the mean disc lifetime decreases as the stellar mass increases, the variance in lifetimes also decreases.
Looking again at the discs with $\alpha=10^{-3}$, for those around low-mass 0.1$\msun$ stars, lifetimes range from 0.12 Myr--20 Myr, roughly two orders of magnitude.\footnote{This variance would be even more pronounced, if the discs were allowed to evolve beyond 20 Myr.}
For more massive stars, the range in disc lifetimes was equal to 0.34 Myr--12 Myr for 0.3$\msun$ stars, and 0.86 Myr--2.9 Myr for 1$\msun$ stars.
The lack of variation in disc lifetimes for more massive stars, shows that the general evolution of those discs in terms of losing mass, doesn't depend greatly on the strength of internal and external photoevaporation within the ranges simulated here.
For less massive stars, the variance in disc lifetimes increases where the effects of internal and external photoevaporation rates are more effective.
When both rates are at their maximum, i.e. the top right of each panel, disc lifetimes are less than 0.3 Myr, whilst when the parameters are set to the minimum, lifetimes are greater than 20 Myr.
Such a large variance in the evolution times of discs could have a significant impact in the types of planetary systems that form within them, as well as yielding surprising observations that are difficult to explain with traditional models (e.g. Peter Pan discs \citep{Silverberg16}).

As the stellar mass increases, the propensity of discs undergoing {\it IO-A} (red region) and {\it OI-G} (light blue region) increases.
With the mass loss rate due to X-rays emanating from the central star increasing, this causes gaps to form quite easily in the disc around the effective gravitational radius.
For discs in stronger external UV environments, as well as those around more X-ray luminous stars, the outer discs are able to be dispersed before the inner disc has accreted onto the central star, increasing the frequency at which the more complex evolution pathways occur.
For Solar-mass stars, panel a) shows the extreme of this, where the most common evolution pathway is {\it IO-A}, as gaps are easily able to form in the discs, but as $\alpha=10^{-4}$, the time-scale for the inner discs to accrete on to the central stars is much longer than the dispersal time-scale of the outer disc region.
As $\alpha$ increases, going from panels a) to c), the viscous time-scale of the inner discs decreases, allowing them to accrete on to the central stars, opening holes in the discs before the outer disc fully disperses, which allows them to be classified as the more traditional {\it IO} pathway.

Another interesting effect that arises when increasing $\alpha$ is that a larger proportion of discs are able to undergo {\it OI} evolution.
This is due to the larger $\alpha$ values leading to larger viscosities that forces gas to flow through the discs at faster rates.
The faster mass-flow rate through the disc stops gaps forming around the effective gravitational radius due to internal photoevaporation, as it replenishes the supply of gas.
Delaying the time at which a gap opens in the disc allows external photoevaporation to further truncate the outer disc, where in some cases, the truncation stops a gap forming entirely, resulting in more discs undergoing {\it OI} evolution.

\subsubsection{Comparing transition discs with our models }
The black dots in fig. \ref{fig:multiplot} represent observed transition discs in the Orion A star forming region \citep{Kim13}.
The majority of these transition discs contain significant gaps in the inner disc, whilst also showing signs of accretion on to their parent stars, indicating the presence of a small inner disc.
In placing them in fig. \ref{fig:multiplot} we grouped the observations together through stellar mass, with tolerances of 0.25 logarithms, i.e. for solar-mass stars $-0.25<\log{M_*}<0.25$.
Note that no stars within the data contained masses consistent with the criteria for 0.1$\msun$ stars.
Interestingly the observed transition discs mainly line in regions where we expect either {\it inside-out (IO)} or {\it inside-out with continued accretion (IO-A)} evolution pathways.
This is expected since these two formation pathways actively contain transition discs for a significant portion of the disc lifetimes.
If some of the observations fell in the {\it outside-in} evolution pathway, then this would be in disagreement with the models, possibly indicating that those models are not adequately capturing the evolution of protoplanetary discs. Finding transition discs in a strong UV environment would hence provide an interesting test of our understanding of the final dispersal of protoplanetary discs. 

\subsection{More massive discs}
\label{sec:disc_mass}

\begin{figure*}
    \centering
    \vspace{-0.3cm}    
    \includegraphics[width=\textwidth]{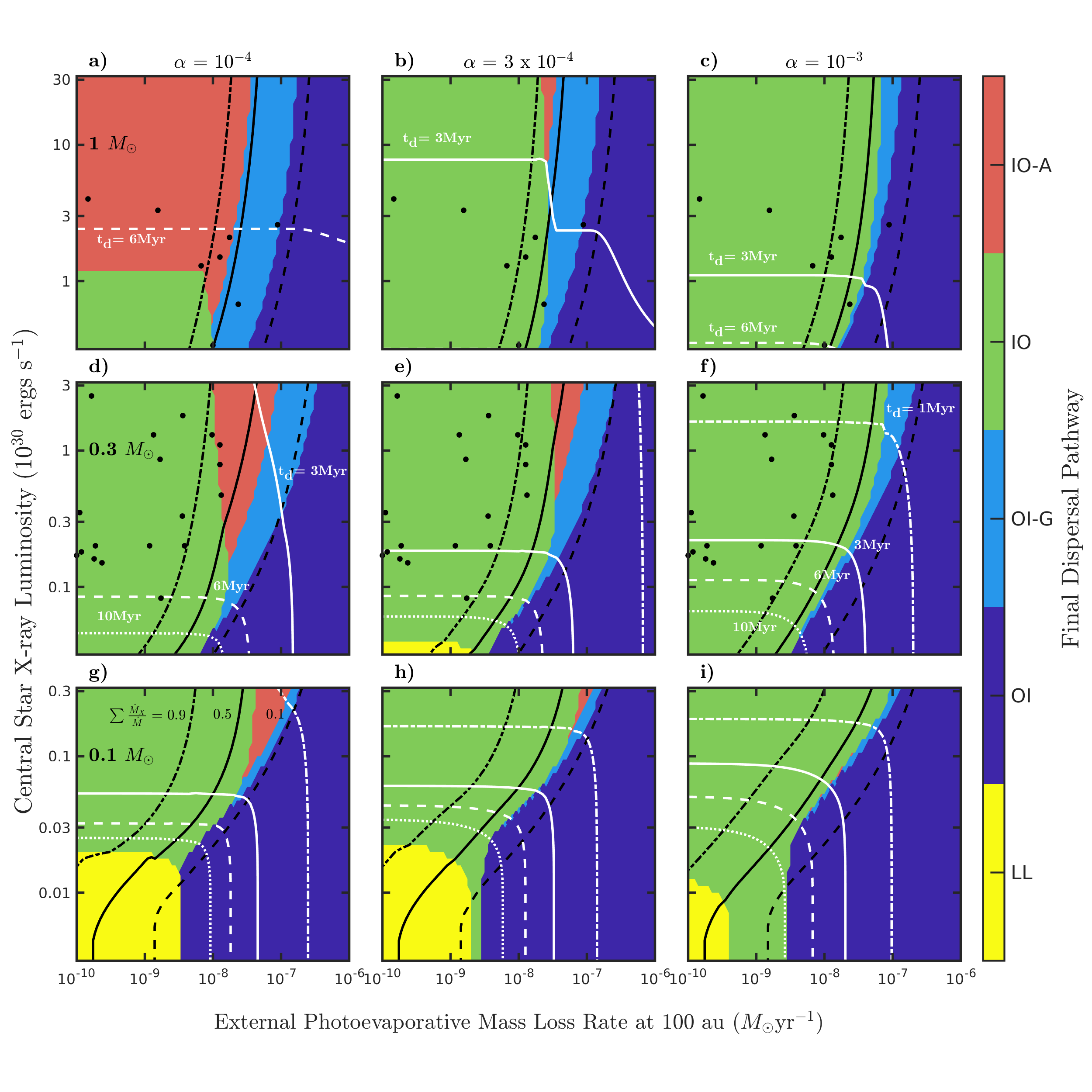}
    \vspace{-1.4cm}
    \caption{Final dispersal types for protoplanetary discs around stars of different masses: 1 $\msun$ (top panels), 0.3 $\msun$ (middle panels), and 0.1 $\msun$ (bottom panels).
    Initial disc masses for these discs was set to the maximum stable mass (eq. \ref{eq:max_disc_mass}).
    The effects of changing the viscosity parameter $\alpha$ is also shown with: $\alpha = 10^{-4}$ (left-hand panels), $\alpha = 3\times 10^{-4}$ (middle panels), and $\alpha = 10^{-3}$ (right-hand panels),
    The final dispersal types are as follows: IO-A = {\it Inside-Out with accreting inner disc} , IO = {\it Inside-Out dispersal}, OI-G = {\it Outside-In with Gaps}, OI = {\it Outside-In} and LL = {\it Long-lived discs}.
    Black lines show the fraction of the mass lost through internal photoevaporation compared to the total mass mass lost: dot-dashed = 0.9, solid = 0.5, dashed = 0.1.
    Lifetimes of the discs are shown by the white lines: dot-dashed = 1 Myr, solid = 3 Myr, dashed = 6 Myr, dotted = 10 Myr.
    We denote the locations of observed transition discs with black points using data from \citet{Kim13}.}
    \label{fig:multiplot_100}
\end{figure*}

The section above examined discs that were initially half the maximum mass before gravitational instability, as per eq. \ref{eq:max_disc_mass}.
We now examine discs that initially contain masses equal to the maximum stable disc mass (i.e. a factor 2 larger).

One might consider that uniformly increasing the disc mass over the whole parameter space would lead to similar results, just at later times. However that is not necessarily the case since the distribution of mass after a given period of evolution would be different, and depending on the magnitude of the mass-flow through the disc, i.e. $\alpha$, then the discs could follow different dispersal pathways as they evolve.

Figure \ref{fig:multiplot_100} shows the evolution pathways for discs with initial masses equal to the maximum stable disc mass.
All colours, lines and notations are the same as those described above in sects. \ref{sec:low_mass_stars} and \ref{sec:stellar_mass}, as well as fig. \ref{fig:multiplot}.
Comparing fig. \ref{fig:multiplot_100} to fig. \ref{fig:multiplot}, the general trends are the same. However, more pronounced for low-mass stars is the presence of {\it long-lived} discs, in regions of the parameter space where the mass loss rates from both internal and external photoevaporation are weak.
Unsurprisingly given the doubling in initial disc mass, disc lifetimes are increased for all discs simulated, allowing some discs around 0.3$\msun$ stars to be classified as being {\it long-lived}.

A major feature that stands out in all of the panels is the reduction of discs following the intermediate evolution pathways: {\it IO-A} and {\it OI-G}.
Conversely there is also an increase in the likelihood of discs being in the {\it IO} or {\it OI} pathways.
The conversion of discs being {\it IO-A} in lower mass discs to being {\it IO} in the higher mass discs is purely due to the increase in disc mass.
As the discs are initially more massive, there is a greater abundance of gas in the outer regions of the disc, which accompanied by low external mass loss rates, results in a longer time-scale required to fully disperse the outer disc once internal photoevaporation has opened a gap in the inner disc region.
With the dispersal time-scale being longer, this therefore gives the inner disc more time to accrete onto the central star, allowing some discs to form large inner holes and follow the {\it IO} pathway.

For the slight increase in discs following the {\it OI} pathway, this is again due to the increase in disc mass.
With more massive discs, the time at which internal photoevaporation is able to open a gap in the inner disc is delayed, as there is a greater abundance of material in the outer disc that flows through the disc towards the star, replenishing the region where mass is mainly lost due to internal photoevaporation and where gaps initially form.
This delay in the time that a gap can form, also gives more time for external photoevaporation to truncate the outer disc.
Even though there is more mass initially in the outer disc, should the delay in the gap-opening time-scale be longer than the externally driven dispersal time-scale, then the discs are able to be truncated to the point where gaps are no longer formed.
This therefore increases the number of discs following the {\it OI} pathway when the initial disc mass is increased.

\section{Discussion}
\label{sec:discussion}

\subsection{Disc fractions over time}

Observations of multiple star-forming regions yield distributions of disc fractions as a function of age \citep[see for example][]{Ribas15}.
These distributions are then normally used to inform upon protoplanetary disc evolution, i.e. by determining an exponential decay time-scale for disc lifetimes. However, it should be noted that a simple disc fraction as a function of cluster age neglects ongoing star formation and the wide spread in actual stellar ages in a region \citep[e.g.][]{2014ApJ...787..108G, 2014ApJ...787..109G}.

To compare our results with observed disc fractions over time, we generate a population of stars of masses between 0.1--1 $\msun$, and then calculate their expected lifetime by interpolating the simulation results the make up fig. \ref{fig:multiplot}.
To generate the populations, we draw stellar masses $M_*$, from a \citet{Kroupa01} initial mass function
\begin{equation}
    \xi (M_*)\propto \left\{ \begin{array}{ll}
M_*^{-1.3} & {\rm for}~ 0.08\msun\le M_* < 0.5\msun \\
\\
M_*^{-2.3} & {\rm for}~ 0.5\msun\le M_* < 10\msun.
\end{array} \right.
\end{equation}
For each star we follow the linear relation found in \citet{Flaischlen21} and define an X-ray luminosity by randomly drawing a value from a normal distribution where the mean is equal to
\begin{equation}
    \log{\overline{L_X}} = 30.58 + 2.08\log{\left(\frac{M_*}{\msun}\right)} 
\end{equation}
with a standard deviation of 0.29.
The means calculated are comparable to the mid-point values in table \ref{tab:parameters} for stars of mass 0.1, 0.3 and 1 $\msun$.
Note that when drawing the stellar mass and X-ray luminosity, we ensure that their values are within the range of simulations presented in sect. \ref{sec:main_plots}, i.e. we do not extrapolate out of our simulated grid.

We then generate a population of 10,000 discs and determine the lifetime of each disc by interpolating within the data comprised in fig. \ref{fig:multiplot}.
With the lifetimes of each population known, we can then calculate the disc fraction as each population ages.
Given that not all stars form at the same time, and instead form over a fraction of a clusters lifetime, we assume that there is an initial burst of star formation at a constant rate, before the rate exponentially decays with a time-scale, $t_{\rm decay}$,
\begin{equation}
    {\rm SFR}\propto {\rm SFR_0}\times\left\{ \begin{array}{ll}
1 & t < 1 {\rm Myr} \\
\\
\exp{\left[\dfrac{-(t-1 {\rm Myr})^2}{t_{\rm decay}^2}\right]} & t \ge 1 {\rm Myr}
\end{array} \right.
\end{equation}
where ${\rm SFR_0}$ is the constant star formation rate during the first million years.
With the prescribed star formation distribution above we then pick a random formation time ($t_{\rm form}$) for each disc, which we then add to the disc lifetime to incorporate into the comparison of disc fraction against age of the cluster.

Note that observed disc fractions depend on which part of the disc the observation was tracing, which is usually the inner $\leq 2$\,au, e.g. as in Spitzer observations. Here we compare with our model disc lifetimes, which may not strictly correlate since we require the removal of the entire disc, making them upper limits on the inferred disc lifetimes. 

\begin{figure}
    \centering
    \includegraphics[scale=0.6]{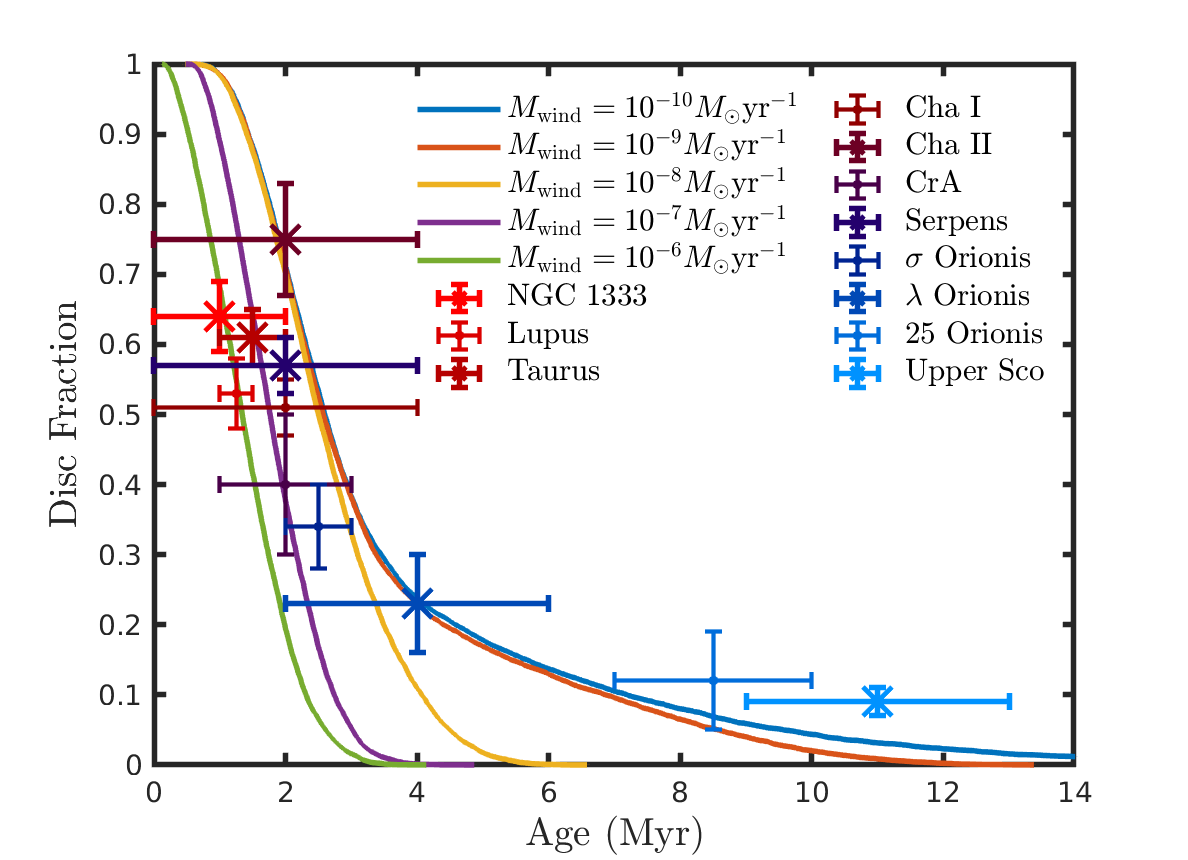}\\
    \caption{Disc fractions as a function of age of a cluster for different simulated populations in different external environments. Simulated lines correspond to populations from discs with a viscous $\alpha=10^{-3}$, and with random formation times using $t_{\rm decay   } = 1$ Myr. Points denote observed disc fractions for a variety of clusters where the fraction is calculated by determining how many stars contain infra-red excesses \citep{Ribas15}. }
    \label{fig:disc_fraction}
\end{figure}

\begin{figure}
    \centering
    \includegraphics[scale=0.6]{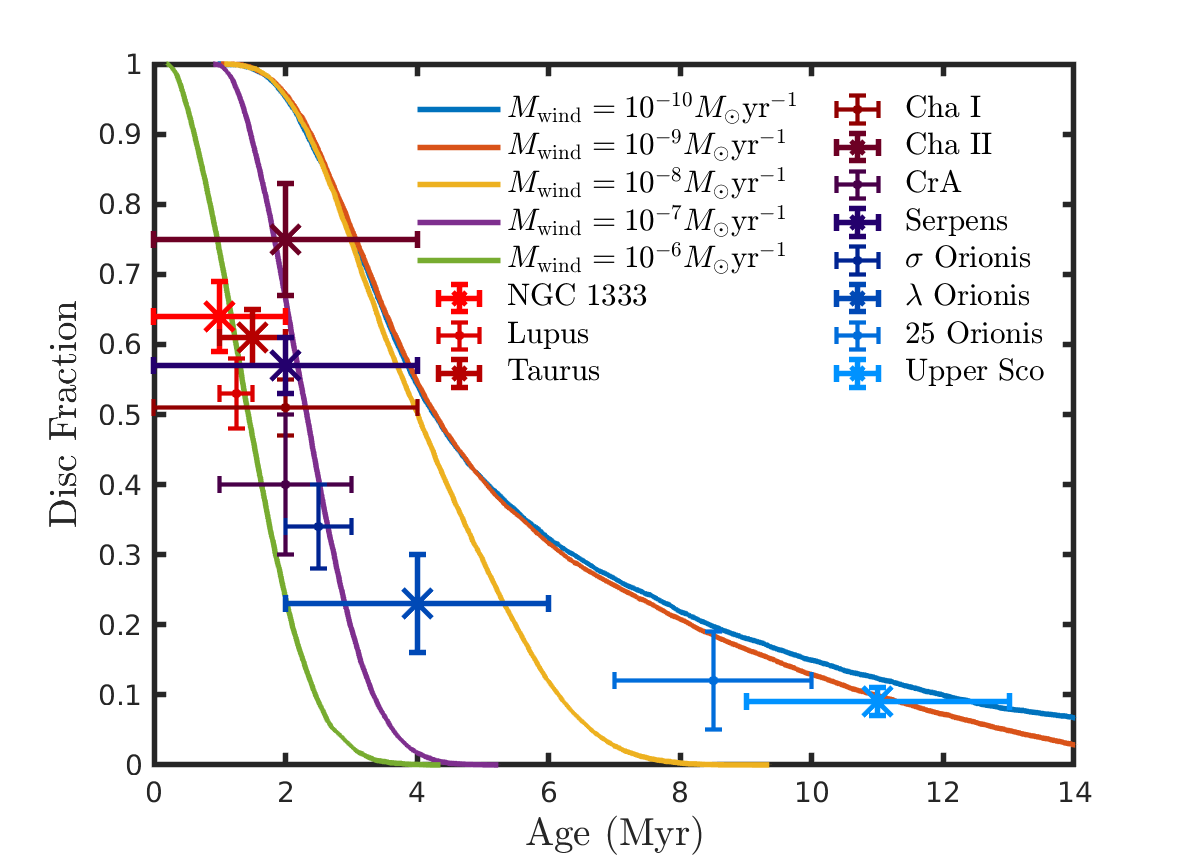}\\
    \caption{Same as fig. \ref{fig:disc_fraction}, but for discs with initial disc masses equal to the maximum stable disc mass (eq. \ref{eq:max_disc_mass}.)}
    \label{fig:disc_fraction_100}
\end{figure}

Figure \ref{fig:disc_fraction} shows the disc fraction against cluster age for protoplanetary discs evolving with different external photoevaporation rates (lines), i.e evolving in different external environments, with a viscous $\alpha = 10^{-3}$.
The points denote observed cluster ages and disc fractions along with their associated errors \citep{Ribas15}.
For the initial 1 Myr of each population, there is little difference in the disc fractions, since photoevaporation has had too little time to significantly disperse enough discs. Shortly after though there is a dramatic fall in disc fraction with time, though it is still the case that this rapid disc destruction is still consistent with the decline in observed disc fractions irrespective of the modelled external photoevaporation rate.

However, by and beyond $\sim$2 Myr the differences become more distinct. At $\sim$2 Myr all discs evolving under the strongest external UV fields are gone. Placing discs in progressively weaker UV environments leads to a progressively shallower tail in disc fraction. By an external mass loss rate normalisation of $10^{-9}\msunyr$, external photoevaporation has little effect on the disc fraction tail off, which is not unexpected, given that from fig. \ref{fig:multiplot}, {\it IO} or {\it IO-A} are the most common evolution pathways for discs with low external mass loss rates.

Beyond 3\,Myr, our models suggest that the low (but non-zero) disc fractions in the tail require low external UV fields/external photoevaporative mass loss rates. Indeed, only external mass-loss rates <$10^{-9}\msunyr$ are able to come close to matching the observed disc fractions in 25 Orionis or Upper Sco, and even then they predict smaller disc fractions than that observed. However, this could again be explained if ongoing low mass star formation is occurring in these regions after the primary burst of star formation (including massive star formation). 

There are other reasons as to why the observed disc fractions could be larger for older clusters than what our models predict, including a weaker average X-ray luminosity than that simulated here, as well as larger initial disc masses.
For fig. \ref{fig:disc_fraction} the initial disc masses was equal to half the maximum stable disc mass (eq. \ref{eq:max_disc_mass}), which as has been seen when comparing the lifetimes of these discs to those more massive (white lines in figs. \ref{fig:multiplot} and \ref{fig:multiplot_100}), leads to shorter average lifetimes.
In initialising the discs with masses equal to the maximum stable disc mass, fig. \ref{fig:disc_fraction_100} indeed shows that the observed disc fractions in older clusters can be reproduced from our models containing larger initial disc masses.

\begin{figure}
    \centering
    \includegraphics[scale=0.6]{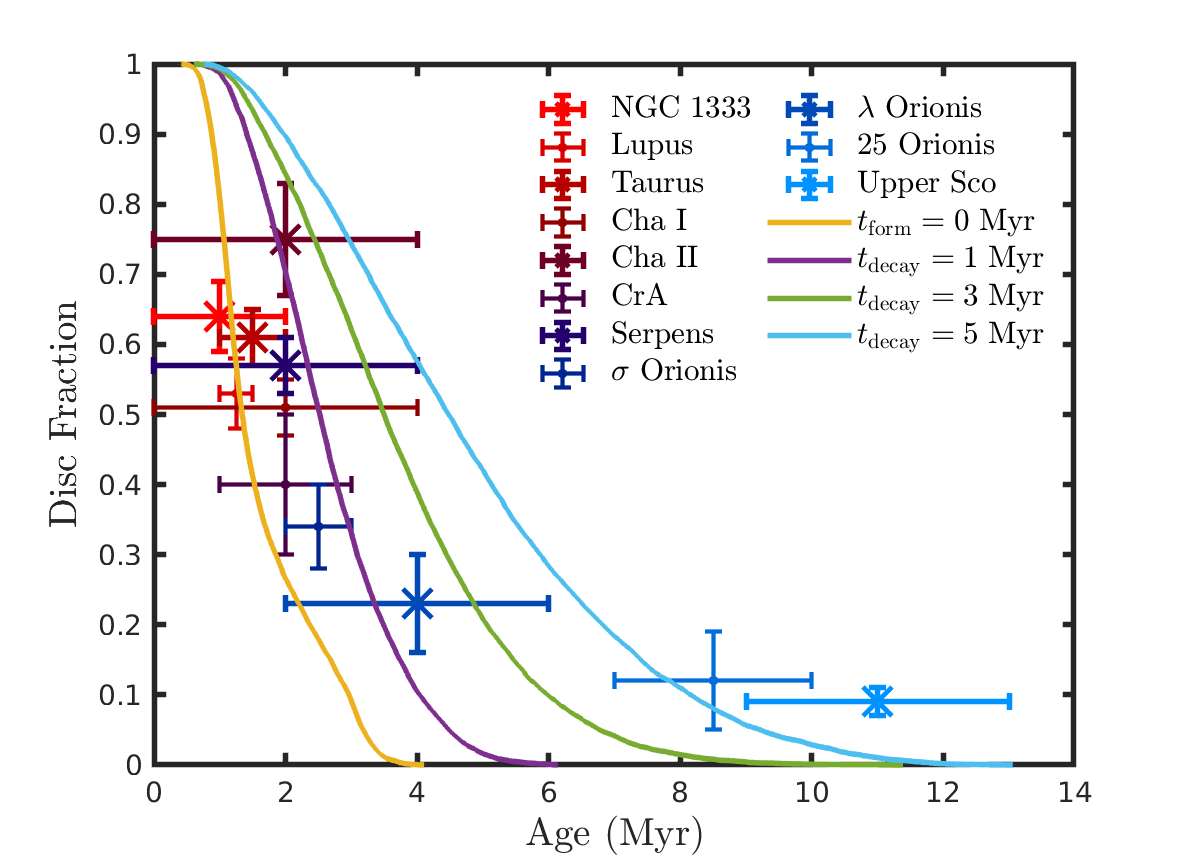}\\
    \caption{Disc fractions as a function of age of a cluster for different simulated populations with random formation times: no formation time (yellow line), exponential decay time $t_{\rm decay}=1$ Myr (purple line), exponential decay time $t_{\rm decay}=3$ Myr (green line), and exponential decay time $t_{\rm decay}=5$ Myr (cyan line). Simulated lines correspond to populations from discs with a viscous $\alpha=10^{-3}$, and with an initial external mass loss rate of $10^{-8}\msunyr$. Points denote observed disc fractions for a variety of clusters where the fraction is calculated by determining how many stars contain infra-red excesses \citep{Ribas15}.}
    \label{fig:disc_frac_tform}
\end{figure}

Whilst fig. \ref{fig:disc_fraction} showed disc fractions in young clusters (<3 Myr) could be reproduced with a range of different external mass-loss rates, there is also an added layer of degeneracy.
This involves the star formation rate (SFR) and how this rate changes over the lifetime of a cluster \citep[e.g.][]{2014ApJ...787..108G, 2014ApJ...787..109G}.
Though other works may not include a delay in the formation of some stars, fig. \ref{fig:disc_fraction} assumes a constant SFR over the first 1 Myr, followed by an exponential decay in SFR with a decay time-scale of 3 Myr, thus simulating ongoing star formation seen in large-scale cluster models.
Figure \ref{fig:disc_frac_tform} shows the evolving disc fractions for discs with external mass-loss rates equal to $10^{-8}\msunyr$, incorporating different time-scales for star formation and decay time-scales: 0 Myr formation time with no exponential decay (yellow), an exponential decay time-scale of 1 Myr (purple), a decay time-scale of 3 Myr (green) and a decay-timescale of 5 Myr (cyan).
As expected with star formation occurring over longer periods, disc fractions remain larger for a longer period of time.
Even including a decay time-scale of 1 Myr, leads to a noticeable difference compared to all discs forming at the same time, i.e. 0 Myr.
From fig. \ref{fig:disc_frac_tform} it is clear that the range of disc fractions when including ongoing star formation with longer decay time-scales, i.e. comparing the yellow line through to the cyan line, adds a layer degeneracy to determining the local environment, and the disc evolution parameters, that observed clusters lie in.
This is due to the overlap in the range of star formation time-scales with the range of external mass-loss rates seen in fig. \ref{fig:disc_fraction}.

\subsection{Peter Pan discs}
Recently protoplanetary discs around stars with ages >40 Myr have been observed \citep{Silverberg16,Silverberg20,LeeSongMurphy20}.
The presence of such discs, dubbed ``Peter Pan Discs'' was a surprise because they were so much older than expected from disc fractions \citep[e.g.][]{Ribas15} and prior theoretical models.
Numerical models by \citet{Coleman20} showed that such discs could exist should they form and evolve either on the periphery of low-mass star forming regions, or deeply embedded for a significant fraction of their lifetime (i.e. exposed to very weak UV environments).
Whilst \citet{Coleman20} only examined such discs around 0.1$\msun$ stars, \citet{wilhelm22} expanded on those models to examine whether Peter Pan discs could form around more massive stars, finding that they could indeed form around stars with masses $\le0.6 \msun$, with more massive stars unable to host such discs due to inner gaps forming, or too efficient accretion and photoevaporation from the host star.

From figs. \ref{fig:multiplot} and \ref{fig:multiplot_100} we find that Peter Pan discs can form around low mass stars with low external mass-loss rates consistent with previous works \citep{Coleman20,wilhelm22}.
An added constraint that our results provide is that the X-ray luminosity of the star also has to be low compared to the average found in \citet{Flaischlen21}.
Future observations of the X-ray luminosities of identified Peter Pan discs could further lock down the properties of the discs and the environments they inhabited.
Our models are also able to produce some long-lived discs around 0.3$\msun$ stars, but only for more massive discs.
This is in slight agreement with the work of \citet{wilhelm22}, but there is a greater restriction in the parameter space in our models for where such discs form, which will only become further restricted with increasing stellar mass.
This further tightening of restrictions is due to increasing X-ray luminosities with stellar mass, reducing the lifetimes of discs in similar environments, yielding different results to those found in \citet{wilhelm22} for more massive stars.

\subsubsection{Relic discs}
Long-lived discs have also been observed in disc evolution models including only internal photoevaporation \citep[e.g.][]{Owen12}.
Such discs are the remnants of photoevaporated discs around low X-ray luminosity sources, where viscous spreading allowed the discs to spread to large radii, such that X-ray photons were not able to launch effective winds. These discs were considered to be an explanation for a fraction of young debris discs based on infra-red excesses at 24 and 70 $\mu$m \citep[e.g.][]{Wyatt08}. Whilst these discs are consistent with the Peter Pan discs described above, and are in agreement with the need for low stellar X-ray luminosities, this work provides the extra constraint of being in an isolated environment or at least in regions of low external UV fields. This means that where in previous work, relic discs would appear around low X-ray sources, in this work such discs would have much shorter lifetimes if there were in strong external UV environments, and as such would not be classified as relic discs. This again shows the importance of including both internal and external photoevaporation within models concerning protoplanetary disc evolution.

\subsection{Identifying external photoevaporation in distant clusters using the lack transition discs in high UV environments}
Our models suggest that discs dispersing by \textit{OI} should generally not result in transition discs and this is supported by the distribution of \cite{Kim13} transition discs being predominantly in the \textit{IO} regime. This does not entirely preclude the existence of transition discs in the high UV environments of massive star forming regions since stellar clusters dynamically evolve, continuously introducing new discs into the high UV regions and discs can also be shielded by natal star forming gas for some time \citep[e.g.][]{2019MNRAS.490.5478W, 2021arXiv210107826C, Qiao22}. Some discs may therefore have evolved in a weaker UV environment for some time where \textit{IO} would dominate, before being introduced to the high UV environment where \textit{OI} dominates. 

This has two implications. The first is that this is a testable prediction using transition disc fractions in the inner ONC compared to other regions. If validated, the second implication is that this could provide a means of indirectly diagnosing external photoevaporation in distant stellar clusters. External photoevaporation in massive distant clusters such as Carina is expected to be widespread, but observationally verifying this is challenging. This is primarily because they are too distant for cometary proplyds \citep[the usual unambiguous way of identify external photoevaporative winds][]{1994ApJ...436..194O,  2016ApJ...826L..15K, 2021MNRAS.501.3502H} to be spatially resolved. But there is also the added complication that known spatially unresolved wind diagnostics cannot currently separate out contributions from the inner and outer photoevaporative winds (e.g. Ballabio et al. in preparation). 

A statistical under-abundance of transition discs in high UV environments may offer a means of indirectly inferring the widespread impact of external photoevaporation in massive stellar clusters more distant than those at $\sim400\,$pc in Orion. Given that Tr14 and Tr16 in Carina appear to be dynamically young and with high UV fields \citep{2019MNRAS.486.4354R} the only way transition discs should appear is if discs are embedded and hence shielded from the UV radiation field for sufficient time. 

\subsection{Model limitations}
In this paper we have introduced disc models that freely evolve under the impact of plausible internal and external photoevaporative winds as well as viscous transport. Here we summarise some limitations of these calculations.

First, we assume that both the radiation environment (which sets the external photoevaporative mass loss rate) and X-ray luminosity are constant over time for any given disc. In reality, stars move around in clusters leading to a time varying UV environment \citep{Qiao22}.  This may lead to some small degree of cloud shielding before discs enter the high UV environment \citep{Qiao22} which may enable internal photoevaporation to play a greater role when both internal and external winds are significant. The inner X-ray luminosity is known to be variable on short time-scales, but only to an extent that is minor compared to the observed spread in X-ray luminosities \citep{2005ApJS..160..401P, 2017RSOS....470114E}.

Another limitation is that the stellar mass dependent internal X-ray photoevaporation prescription of \citet{Picogna21} is only available until the inner hole opens. Beyond that point the prescription is only for a 0.7\,M$_\odot$ star \citep{Picogna19}. However, we deem it sufficient since that latter prescription was developed considering X-ray luminosities spanning around 3 orders of magnitude and the evolution of the disc is relatively quick once an inner hole opens.

Within the models presented we have only tested a single distribution of mass for the initial disc profiles. Whilst we use an exponent for the initial surface density slope of -1, other works range from -0.5 to -1.5 \citep[e.g. MMSN model][]{Hayashi}. Given that using different slopes will change the distribution of mass across the discs, this could affect the outcomes of the model. However we would only expect this to affect the locations of the transitions between the different evolutionary pathways, since in the more extreme regions of the parameter space (e.g. the top-left or bottom-right of the panels in figs. \ref{fig:multiplot} and \ref{fig:multiplot_100}), internal or external photoevaporation should continue to dominate, with the only different being the disc lifetimes which will depend on where the mass is located. In regards to the transition regimes, {\it IO-A} and {\it OI-G}, their prevalence could be expected to increase for more shallower profiles since gaps should more easily form and the outer disc should live longer (due to more mass initially located there), whilst they may become rarer for steeper slopes where little mass is situated in the outer disc. Further studies of the effects of the initial conditions will be left to future work which could then be better compared with observations that could give clues about the initial conditions for observed protoplanetary discs

Finally, and probably of most significance, is that our models do not include any prescription for internal magnetohydrodynamically (MHD) driven winds. Internal winds can also be driven magnetohydrodynamically, through the magnetorotational instability \citep[MRI,][]{1991ApJ...376..214B, 2009ApJ...691L..49S, 2016A&A...596A..74S} or magnetocentrifugal winds \citep{2007prpl.conf..277P, 2014prpl.conf..411T, 2016ApJ...818..152B, 2020ApJ...896..126G}. It is anticipated the MHD winds lead to comparable mass loss rates to internal photoevaporation and control the rate of material transport if discs are low viscosity. \citep[e.g.][]{Hasegawa22, 2021MNRAS.tmpL.109T}. A future improvement on this work would be to include a treatment of the MHD winds, but in the absence of that we suggest that our high internal photoevaporation, high viscosity discs probably represent a proxy of MHD driven dispersal and material transport in low viscosity discs.

\section{Summary and Conclusions}
\label{sec:conclusions}

In this work we have explored the different evolution pathways for protoplanetary discs undergoing both internal and external photoevaporation. We have done so for a parameter space spanning a range of stellar masses, internal X-ray luminosities representative of the observed spread, external photoevaporative mass loss rates representative of low mass star forming regions through to high UV environments and three different degrees of viscous transport. Unlike previous work we have placed no constraints on the degree of internal/external clearing to fine tune disc lifetimes to be similar to the instantaneous disc fractions in stellar clusters.  We draw the following main conclusions from this work. \\

1. We identify five different evolutionary/clearing pathways for discs. i) {\it long-lived} discs that survive for $>20$\,Myr. ii) Discs cleared from the {\it inside-out}, where internal X-ray/EUV photoevaporation dominates and creates an increasingly large hole. iii) {\it outside-in} dispersal, where external photoevaporation truncates the disc and the remaining disc at small radii, the very inner disc, disappears viscously. There are then two intermediate cases with lingering inner material where the internal and external radiation fields both have a significant effect on the disc. We identify which of the two by empirically choosing a ``clearing radius'' in the disc that gives a good distinction depending on whether an inner hole or outer truncation reaches it first. The intermediate cases are iv) {\it inside-out with continued accretion}, where there is a largening hole but a compact (<1\,au) region that continues accreting and v) {\it outside-in with gap features} where the disc is truncated but a gap is also opened up between the very inner and outer disc for a period of time. For a schematic of these evolutionary pathways see fig. \ref{fig:evolution_cartoon}, whilst table \ref{tab:dispersal_types} outlines the criteria and descriptions for each evolutionary pathway. \\
 
2. We find that the evolutionary time-scale of the inner $\sim2$\,au is comparable to or greater than the time-scale for dispersal of the outer disc in high UV environments. We therefore do not expect transition discs in high UV environments, or at the least expect an under-abundance of them (e.g. they may exist if the disc spent some time shielded). This appears to be in agreement with the locations of transition discs towards the Orion Nebular Cluster. An under-abundance of transition discs in distant clusters may therefore provide a means of diagnosing external photoevaporation. \\

3. Randomly sampling our models, we find that we require ongoing star formation to reproduce observed disc fractions in clusters with time. We also find that the impact of environment on that function really only manifests in the first $\sim 3$\,Myr, after which the slow decline in disc fraction is probably set by weaker levels of ongoing star formation.   \\

4. Our models are compatible with the rare occurrence of very long lived ($>40$\,Myr) ``Peter Pan'' discs. This is consistent with previous models, but with the added constraint introduced based on the possible internal X-ray luminosity. As one would expect, Peter Pan discs require both low external photoevaporation rates (and hence weak external FUV environments) and low internal X-ray luminosities to survive for so long.   \\

Overall this work demonstrates the importance of a more holistic approach to improving our understanding of protoplanetary disc evolution and dispersal.
It is clear that including both internal and external photoevaporation mechanisms have important roles in determining the lifetimes and evolution pathways of protoplanetary discs.
This could also have important consequences on the types of planets and planetary systems that form in such discs, which will be studied in future work.

\section*{Data Availability}
The data underlying this article will be shared on reasonable request to the corresponding author.

\section*{Acknowledgements}
The authors thank James Owen for providing useful and interesting comments that improved the paper.
GALC was funded by the Leverhulme Trust through grant RPG-2018-418.
TJH is funded by a Royal Society Dorothy Hodgkin Fellowship.
This research utilised Queen Mary's Apocrita HPC facility, supported by QMUL Research-IT (http://doi.org/10.5281/zenodo.438045).

\vspace{-0.2cm}
\bibliographystyle{mnras}
\bibliography{references}{}

\begin{thebibliography}{}
\makeatletter
\relax
\def\mn@urlcharsother{\let\do\@makeother \do\$\do\&\do\#\do\^\do\_\do\%\do\~}
\def\mn@doi{\begingroup\mn@urlcharsother \@ifnextchar [ {\mn@doi@}
  {\mn@doi@[]}}
\def\mn@doi@[#1]#2{\def\@tempa{#1}\ifx\@tempa\@empty \href
  {http://dx.doi.org/#2} {doi:#2}\else \href {http://dx.doi.org/#2} {#1}\fi
  \endgroup}
\def\mn@eprint#1#2{\mn@eprint@#1:#2::\@nil}
\def\mn@eprint@arXiv#1{\href {http://arxiv.org/abs/#1} {{\tt arXiv:#1}}}
\def\mn@eprint@dblp#1{\href {http://dblp.uni-trier.de/rec/bibtex/#1.xml}
  {dblp:#1}}
\def\mn@eprint@#1:#2:#3:#4\@nil{\def\@tempa {#1}\def\@tempb {#2}\def\@tempc
  {#3}\ifx \@tempc \@empty \let \@tempc \@tempb \let \@tempb \@tempa \fi \ifx
  \@tempb \@empty \def\@tempb {arXiv}\fi \@ifundefined
  {mn@eprint@\@tempb}{\@tempb:\@tempc}{\expandafter \expandafter \csname
  mn@eprint@\@tempb\endcsname \expandafter{\@tempc}}}

\bibitem[\protect\citeauthoryear{{Adams}, {Hollenbach}, {Laughlin}  \&
  {Gorti}}{{Adams} et~al.}{2004}]{2004ApJ...611..360A}
{Adams} F.~C.,  {Hollenbach} D.,  {Laughlin} G.,   {Gorti} U.,  2004, \mn@doi
  [\apj] {10.1086/421989}, \href
  {https://ui.adsabs.harvard.edu/abs/2004ApJ...611..360A} {611, 360}

\bibitem[\protect\citeauthoryear{{Alarc{\'o}n} et~al.,}{{Alarc{\'o}n}
  et~al.}{2021}]{2021ApJS..257....8A}
{Alarc{\'o}n} F.,  et~al., 2021, \mn@doi [\apjs] {10.3847/1538-4365/ac22ae},
  \href {https://ui.adsabs.harvard.edu/abs/2021ApJS..257....8A} {257, 8}

\bibitem[\protect\citeauthoryear{{Alexander} \& {Armitage}}{{Alexander} \&
  {Armitage}}{2007}]{Alexander07}
{Alexander} R.~D.,  {Armitage} P.~J.,  2007, \mn@doi [\mnras]
  {10.1111/j.1365-2966.2006.11341.x}, \href
  {http://adsabs.harvard.edu/abs/2007MNRAS.375..500A} {375, 500}

\bibitem[\protect\citeauthoryear{{Alexander} \& {Armitage}}{{Alexander} \&
  {Armitage}}{2009}]{Alexander09}
{Alexander} R.~D.,  {Armitage} P.~J.,  2009, \mn@doi [\apj]
  {10.1088/0004-637X/704/2/989}, \href
  {http://adsabs.harvard.edu/abs/2009ApJ...704..989A} {704, 989}

\bibitem[\protect\citeauthoryear{{Alexander} \& {Pascucci}}{{Alexander} \&
  {Pascucci}}{2012}]{AlexanderPascucci12}
{Alexander} R.~D.,  {Pascucci} I.,  2012, \mn@doi [\mnras]
  {10.1111/j.1745-3933.2012.01243.x}, \href
  {http://adsabs.harvard.edu/abs/2012MNRAS.422L..82A} {422, 82}

\bibitem[\protect\citeauthoryear{{Alexander}, {Clarke}  \&
  {Pringle}}{{Alexander} et~al.}{2005}]{Alexander05}
{Alexander} R.~D.,  {Clarke} C.~J.,   {Pringle} J.~E.,  2005, \mn@doi [\mnras]
  {10.1111/j.1365-2966.2005.08786.x}, \href
  {https://ui.adsabs.harvard.edu/abs/2005MNRAS.358..283A} {358, 283}

\bibitem[\protect\citeauthoryear{{Andrews}, {Wilner}, {Hughes}, {Qi}  \&
  {Dullemond}}{{Andrews} et~al.}{2010}]{Andrews10}
{Andrews} S.~M.,  {Wilner} D.~J.,  {Hughes} A.~M.,  {Qi} C.,   {Dullemond}
  C.~P.,  2010, \mn@doi [\apj] {10.1088/0004-637X/723/2/1241}, \href
  {https://ui.adsabs.harvard.edu/abs/2010ApJ...723.1241A} {723, 1241}

\bibitem[\protect\citeauthoryear{{Andrews} et~al.,}{{Andrews}
  et~al.}{2018}]{2018ApJ...869L..41A}
{Andrews} S.~M.,  et~al., 2018, \mn@doi [\apjl] {10.3847/2041-8213/aaf741},
  \href {https://ui.adsabs.harvard.edu/abs/2018ApJ...869L..41A} {869, L41}

\bibitem[\protect\citeauthoryear{{Ansdell}, {Williams}, {Manara}, {Miotello},
  {Facchini}, {van der Marel}, {Testi}  \& {van Dishoeck}}{{Ansdell}
  et~al.}{2017}]{2017AJ....153..240A}
{Ansdell} M.,  {Williams} J.~P.,  {Manara} C.~F.,  {Miotello} A.,  {Facchini}
  S.,  {van der Marel} N.,  {Testi} L.,   {van Dishoeck} E.~F.,  2017, \mn@doi
  [\aj] {10.3847/1538-3881/aa69c0}, \href
  {http://adsabs.harvard.edu/abs/2017AJ....153..240A} {153, 240}

\bibitem[\protect\citeauthoryear{{Ansdell} et~al.,}{{Ansdell}
  et~al.}{2018}]{2018ApJ...859...21A}
{Ansdell} M.,  et~al., 2018, \mn@doi [\apj] {10.3847/1538-4357/aab890}, \href
  {https://ui.adsabs.harvard.edu/abs/2018ApJ...859...21A} {859, 21}

\bibitem[\protect\citeauthoryear{{Bai}, {Ye}, {Goodman}  \& {Yuan}}{{Bai}
  et~al.}{2016}]{2016ApJ...818..152B}
{Bai} X.-N.,  {Ye} J.,  {Goodman} J.,   {Yuan} F.,  2016, \mn@doi [apj]
  {10.3847/0004-637X/818/2/152}, \href
  {https://ui.adsabs.harvard.edu/abs/2016ApJ...818..152B} {818, 152}

\bibitem[\protect\citeauthoryear{{Balbus} \& {Hawley}}{{Balbus} \&
  {Hawley}}{1991}]{1991ApJ...376..214B}
{Balbus} S.~A.,  {Hawley} J.~F.,  1991, \mn@doi [\apj] {10.1086/170270}, \href
  {https://ui.adsabs.harvard.edu/abs/1991ApJ...376..214B} {376, 214}

\bibitem[\protect\citeauthoryear{{Birnstiel}, {Klahr}  \&
  {Ercolano}}{{Birnstiel} et~al.}{2012}]{2012A&A...539A.148B}
{Birnstiel} T.,  {Klahr} H.,   {Ercolano} B.,  2012, \mn@doi [\aap]
  {10.1051/0004-6361/201118136}, \href
  {https://ui.adsabs.harvard.edu/abs/2012A&A...539A.148B} {539, A148}

\bibitem[\protect\citeauthoryear{{Clarke}, {Gendrin}  \& {Sotomayor}}{{Clarke}
  et~al.}{2001}]{Clarke2001}
{Clarke} C.~J.,  {Gendrin} A.,   {Sotomayor} M.,  2001, \mn@doi [\mnras]
  {10.1046/j.1365-8711.2001.04891.x}, \href
  {http://adsabs.harvard.edu/abs/2001MNRAS.328..485C} {328, 485}

\bibitem[\protect\citeauthoryear{{Cleeves}, {{\"O}berg}, {Wilner}, {Huang},
  {Loomis}, {Andrews}  \& {Czekala}}{{Cleeves}
  et~al.}{2016}]{2016ApJ...832..110C}
{Cleeves} L.~I.,  {{\"O}berg} K.~I.,  {Wilner} D.~J.,  {Huang} J.,  {Loomis}
  R.~A.,  {Andrews} S.~M.,   {Czekala} I.,  2016, \mn@doi [\apj]
  {10.3847/0004-637X/832/2/110}, \href
  {http://adsabs.harvard.edu/abs/2016ApJ...832..110C} {832, 110}

\bibitem[\protect\citeauthoryear{{Coleman}}{{Coleman}}{2021}]{Coleman21}
{Coleman} G. A.~L.,  2021, \mn@doi [\mnras] {10.1093/mnras/stab1904}, \href
  {https://ui.adsabs.harvard.edu/abs/2021MNRAS.506.3596C} {506, 3596}

\bibitem[\protect\citeauthoryear{{Coleman} \& {Haworth}}{{Coleman} \&
  {Haworth}}{2020}]{Coleman20}
{Coleman} G. A.~L.,  {Haworth} T.~J.,  2020, \mn@doi [\mnras]
  {10.1093/mnrasl/slaa098}, \href
  {https://ui.adsabs.harvard.edu/abs/2020MNRAS.496L.111C} {496, L111}

\bibitem[\protect\citeauthoryear{{Coleman} \& {Nelson}}{{Coleman} \&
  {Nelson}}{2014}]{ColemanNelson14}
{Coleman} G.~A.~L.,  {Nelson} R.~P.,  2014, \mn@doi [\mnras]
  {10.1093/mnras/stu1715}, \href
  {http://adsabs.harvard.edu/abs/2014MNRAS.445..479C} {445, 479}

\bibitem[\protect\citeauthoryear{{Coleman} \& {Nelson}}{{Coleman} \&
  {Nelson}}{2016}]{ColemanNelson16}
{Coleman} G.~A.~L.,  {Nelson} R.~P.,  2016, \mn@doi [\mnras]
  {10.1093/mnras/stw149}, \href
  {http://adsabs.harvard.edu/abs/2016MNRAS.457.2480C} {457, 2480}

\bibitem[\protect\citeauthoryear{{Concha-Ram{\'\i}rez}, {Wilhelm}, {Portegies
  Zwart}  \& {Haworth}}{{Concha-Ram{\'\i}rez}
  et~al.}{2019}]{2019MNRAS.490.5678C}
{Concha-Ram{\'\i}rez} F.,  {Wilhelm} M. J.~C.,  {Portegies Zwart} S.,
  {Haworth} T.~J.,  2019, \mn@doi [\mnras] {10.1093/mnras/stz2973}, \href
  {https://ui.adsabs.harvard.edu/abs/2019MNRAS.490.5678C} {490, 5678}

\bibitem[\protect\citeauthoryear{{Concha-Ram{\'\i}rez}, {Portegies Zwart}  \&
  {Wilhelm}}{{Concha-Ram{\'\i}rez} et~al.}{2021}]{2021arXiv210107826C}
{Concha-Ram{\'\i}rez} F.,  {Portegies Zwart} S.,   {Wilhelm} M. J.~C.,  2021,
  arXiv e-prints, \href {https://ui.adsabs.harvard.edu/abs/2021arXiv210107826C}
  {p. arXiv:2101.07826}

\bibitem[\protect\citeauthoryear{{Dullemond}, {Hollenbach}, {Kamp}  \&
  {D'Alessio}}{{Dullemond} et~al.}{2007}]{Dullemond}
{Dullemond} C.~P.,  {Hollenbach} D.,  {Kamp} I.,   {D'Alessio} P.,  2007,
  Protostars and Planets V, \href
  {http://adsabs.harvard.edu/abs/2007prpl.conf..555D} {pp 555--572}

\bibitem[\protect\citeauthoryear{{Eisner} et~al.,}{{Eisner}
  et~al.}{2018}]{2018ApJ...860...77E}
{Eisner} J.~A.,  et~al., 2018, \mn@doi [\apj] {10.3847/1538-4357/aac3e2}, \href
  {http://adsabs.harvard.edu/abs/2018ApJ...860...77E} {860, 77}

\bibitem[\protect\citeauthoryear{{Eistrup}, {Walsh}  \& {van
  Dishoeck}}{{Eistrup} et~al.}{2018}]{2018A&A...613A..14E}
{Eistrup} C.,  {Walsh} C.,   {van Dishoeck} E.~F.,  2018, \mn@doi [\aap]
  {10.1051/0004-6361/201731302}, \href
  {https://ui.adsabs.harvard.edu/abs/2018A&A...613A..14E} {613, A14}

\bibitem[\protect\citeauthoryear{{Ercolano} \& {Pascucci}}{{Ercolano} \&
  {Pascucci}}{2017}]{2017RSOS....470114E}
{Ercolano} B.,  {Pascucci} I.,  2017, \mn@doi [Royal Society Open Science]
  {10.1098/rsos.170114}, \href
  {https://ui.adsabs.harvard.edu/abs/2017RSOS....470114E} {4, 170114}

\bibitem[\protect\citeauthoryear{{Ercolano}, {Picogna}, {Monsch}, {Drake}  \&
  {Preibisch}}{{Ercolano} et~al.}{2021}]{Ercolano21}
{Ercolano} B.,  {Picogna} G.,  {Monsch} K.,  {Drake} J.~J.,   {Preibisch} T.,
  2021, \mn@doi [\mnras] {10.1093/mnras/stab2590}, \href
  {https://ui.adsabs.harvard.edu/abs/2021MNRAS.508.1675E} {508, 1675}

\bibitem[\protect\citeauthoryear{{Facchini}, {Birnstiel}, {Bruderer}  \& {van
  Dishoeck}}{{Facchini} et~al.}{2017}]{2017A&A...605A..16F}
{Facchini} S.,  {Birnstiel} T.,  {Bruderer} S.,   {van Dishoeck} E.~F.,  2017,
  \mn@doi [\aap] {10.1051/0004-6361/201630329}, \href
  {https://ui.adsabs.harvard.edu/abs/2017A&A...605A..16F} {605, A16}

\bibitem[\protect\citeauthoryear{{Fedele}, {van den Ancker}, {Henning},
  {Jayawardhana}  \& {Oliveira}}{{Fedele} et~al.}{2010}]{Fedele10}
{Fedele} D.,  {van den Ancker} M.~E.,  {Henning} T.,  {Jayawardhana} R.,
  {Oliveira} J.~M.,  2010, \mn@doi [\aap] {10.1051/0004-6361/200912810}, \href
  {https://ui.adsabs.harvard.edu/abs/2010A&A...510A..72F} {510, A72}

\bibitem[\protect\citeauthoryear{{Flaischlen}, {Preibisch}, {Manara}  \&
  {Ercolano}}{{Flaischlen} et~al.}{2021}]{Flaischlen21}
{Flaischlen} S.,  {Preibisch} T.,  {Manara} C.~F.,   {Ercolano} B.,  2021,
  \mn@doi [\aap] {10.1051/0004-6361/202039746}, \href
  {https://ui.adsabs.harvard.edu/abs/2021A&A...648A.121F} {648, A121}

\bibitem[\protect\citeauthoryear{{Getman} et~al.,}{{Getman}
  et~al.}{2014a}]{2014ApJ...787..108G}
{Getman} K.~V.,  et~al., 2014a, \mn@doi [\apj] {10.1088/0004-637X/787/2/108},
  \href {https://ui.adsabs.harvard.edu/abs/2014ApJ...787..108G} {787, 108}

\bibitem[\protect\citeauthoryear{{Getman}, {Feigelson}  \& {Kuhn}}{{Getman}
  et~al.}{2014b}]{2014ApJ...787..109G}
{Getman} K.~V.,  {Feigelson} E.~D.,   {Kuhn} M.~A.,  2014b, \mn@doi [\apj]
  {10.1088/0004-637X/787/2/109}, \href
  {https://ui.adsabs.harvard.edu/abs/2014ApJ...787..109G} {787, 109}

\bibitem[\protect\citeauthoryear{{Gorti}, {Dullemond}  \& {Hollenbach}}{{Gorti}
  et~al.}{2009}]{2009ApJ...705.1237G}
{Gorti} U.,  {Dullemond} C.~P.,   {Hollenbach} D.,  2009, \mn@doi [\apj]
  {10.1088/0004-637X/705/2/1237}, \href
  {http://adsabs.harvard.edu/abs/2009ApJ...705.1237G} {705, 1237}

\bibitem[\protect\citeauthoryear{{Gorti}, {Hollenbach}  \& {Dullemond}}{{Gorti}
  et~al.}{2015}]{2015ApJ...804...29G}
{Gorti} U.,  {Hollenbach} D.,   {Dullemond} C.~P.,  2015, \mn@doi [\apj]
  {10.1088/0004-637X/804/1/29}, \href
  {http://adsabs.harvard.edu/abs/2015ApJ...804...29G} {804, 29}

\bibitem[\protect\citeauthoryear{{Gressel}, {Ramsey}, {Brinch}, {Nelson},
  {Turner}  \& {Bruderer}}{{Gressel} et~al.}{2020}]{2020ApJ...896..126G}
{Gressel} O.,  {Ramsey} J.~P.,  {Brinch} C.,  {Nelson} R.~P.,  {Turner} N.~J.,
   {Bruderer} S.,  2020, \mn@doi [\apj] {10.3847/1538-4357/ab91b7}, \href
  {https://ui.adsabs.harvard.edu/abs/2020ApJ...896..126G} {896, 126}

\bibitem[\protect\citeauthoryear{{Guarcello} et~al.,}{{Guarcello}
  et~al.}{2016}]{2016arXiv160501773G}
{Guarcello} M.~G.,  et~al., 2016, arXiv e-prints, \href
  {https://ui.adsabs.harvard.edu/abs/2016arXiv160501773G} {p. arXiv:1605.01773}

\bibitem[\protect\citeauthoryear{{Haisch}, {Lada}  \& {Lada}}{{Haisch}
  et~al.}{2001}]{Haisch01}
{Haisch} Karl~E. J.,  {Lada} E.~A.,   {Lada} C.~J.,  2001, \mn@doi [\apjl]
  {10.1086/320685}, \href
  {https://ui.adsabs.harvard.edu/abs/2001ApJ...553L.153H} {553, L153}

\bibitem[\protect\citeauthoryear{{Hasegawa} et~al.,}{{Hasegawa}
  et~al.}{2022}]{Hasegawa22}
{Hasegawa} Y.,  et~al., 2022, \mn@doi [\apjl] {10.3847/2041-8213/ac50aa}, \href
  {https://ui.adsabs.harvard.edu/abs/2022ApJ...926L..23H} {926, L23}

\bibitem[\protect\citeauthoryear{{Haworth} \& {Clarke}}{{Haworth} \&
  {Clarke}}{2019}]{2019MNRAS.485.3895H}
{Haworth} T.~J.,  {Clarke} C.~J.,  2019, \mn@doi [\mnras]
  {10.1093/mnras/stz706}, \href
  {https://ui.adsabs.harvard.edu/abs/2019MNRAS.485.3895H} {485, 3895}

\bibitem[\protect\citeauthoryear{{Haworth}, {Facchini}, {Clarke}  \&
  {Cleeves}}{{Haworth} et~al.}{2017}]{2017MNRAS.468L.108H}
{Haworth} T.~J.,  {Facchini} S.,  {Clarke} C.~J.,   {Cleeves} L.~I.,  2017,
  \mn@doi [\mnras] {10.1093/mnrasl/slx037}, \href
  {http://adsabs.harvard.edu/abs/2017MNRAS.468L.108H} {468, L108}

\bibitem[\protect\citeauthoryear{{Haworth}, {Clarke}, {Rahman}, {Winter}  \&
  {Facchini}}{{Haworth} et~al.}{2018}]{2018MNRAS.481..452H}
{Haworth} T.~J.,  {Clarke} C.~J.,  {Rahman} W.,  {Winter} A.~J.,   {Facchini}
  S.,  2018, \mn@doi [\mnras] {10.1093/mnras/sty2323}, \href
  {http://adsabs.harvard.edu/abs/2018MNRAS.481..452H} {481, 452}

\bibitem[\protect\citeauthoryear{{Haworth}, {Cadman}, {Meru}, {Hall},
  {Albertini}, {Forgan}, {Rice}  \& {Owen}}{{Haworth} et~al.}{2020}]{Haworth20}
{Haworth} T.~J.,  {Cadman} J.,  {Meru} F.,  {Hall} C.,  {Albertini} E.,
  {Forgan} D.,  {Rice} K.,   {Owen} J.~E.,  2020, arXiv e-prints, \href
  {https://ui.adsabs.harvard.edu/abs/2020arXiv200106225H} {p. arXiv:2001.06225}

\bibitem[\protect\citeauthoryear{{Haworth}, {Kim}, {Winter}, {Hines}, {Clarke},
  {Sellek}, {Ballabio}  \& {Stapelfeldt}}{{Haworth}
  et~al.}{2021}]{2021MNRAS.501.3502H}
{Haworth} T.~J.,  {Kim} J.~S.,  {Winter} A.~J.,  {Hines} D.~C.,  {Clarke}
  C.~J.,  {Sellek} A.~D.,  {Ballabio} G.,   {Stapelfeldt} K.~R.,  2021, \mn@doi
  [\mnras] {10.1093/mnras/staa3918}, \href
  {https://ui.adsabs.harvard.edu/abs/2021MNRAS.501.3502H} {501, 3502}

\bibitem[\protect\citeauthoryear{{Hayashi}}{{Hayashi}}{1981}]{Hayashi}
{Hayashi} C.,  1981, \mn@doi [Progress of Theoretical Physics Supplement]
  {10.1143/PTPS.70.35}, \href
  {http://adsabs.harvard.edu/abs/1981PThPS..70...35H} {70, 35}

\bibitem[\protect\citeauthoryear{{Hillenbrand}}{{Hillenbrand}}{2005}]{2005astro.ph.11083H}
{Hillenbrand} L.~A.,  2005, arXiv e-prints, \href
  {https://ui.adsabs.harvard.edu/abs/2005astro.ph.11083H} {pp
  astro--ph/0511083}

\bibitem[\protect\citeauthoryear{{Keppler} et~al.,}{{Keppler}
  et~al.}{2018}]{2018A&A...617A..44K}
{Keppler} M.,  et~al., 2018, \mn@doi [\aap] {10.1051/0004-6361/201832957},
  \href {https://ui.adsabs.harvard.edu/abs/2018A&A...617A..44K} {617, A44}

\bibitem[\protect\citeauthoryear{{Kim} et~al.,}{{Kim} et~al.}{2013}]{Kim13}
{Kim} K.~H.,  et~al., 2013, \mn@doi [\apj] {10.1088/0004-637X/769/2/149}, \href
  {https://ui.adsabs.harvard.edu/abs/2013ApJ...769..149K} {769, 149}

\bibitem[\protect\citeauthoryear{{Kim}, {Clarke}, {Fang}  \& {Facchini}}{{Kim}
  et~al.}{2016}]{2016ApJ...826L..15K}
{Kim} J.~S.,  {Clarke} C.~J.,  {Fang} M.,   {Facchini} S.,  2016, \mn@doi
  [\apjl] {10.3847/2041-8205/826/1/L15}, \href
  {https://ui.adsabs.harvard.edu/abs/2016ApJ...826L..15K} {826, L15}

\bibitem[\protect\citeauthoryear{{Komaki}, {Nakatani}  \& {Yoshida}}{{Komaki}
  et~al.}{2021}]{2021ApJ...910...51K}
{Komaki} A.,  {Nakatani} R.,   {Yoshida} N.,  2021, \mn@doi [\apj]
  {10.3847/1538-4357/abe2af}, \href
  {https://ui.adsabs.harvard.edu/abs/2021ApJ...910...51K} {910, 51}

\bibitem[\protect\citeauthoryear{{Kroupa}}{{Kroupa}}{2001}]{Kroupa01}
{Kroupa} P.,  2001, \mn@doi [\mnras] {10.1046/j.1365-8711.2001.04022.x}, \href
  {https://ui.adsabs.harvard.edu/abs/2001MNRAS.322..231K} {322, 231}

\bibitem[\protect\citeauthoryear{{Kunitomo}, {Suzuki}  \&
  {Inutsuka}}{{Kunitomo} et~al.}{2020}]{Kunitomo20}
{Kunitomo} M.,  {Suzuki} T.~K.,   {Inutsuka} S.-i.,  2020, \mn@doi [\mnras]
  {10.1093/mnras/staa087}, \href
  {https://ui.adsabs.harvard.edu/abs/2020MNRAS.492.3849K} {492, 3849}

\bibitem[\protect\citeauthoryear{{Lee}, {Song}  \& {Murphy}}{{Lee}
  et~al.}{2020}]{LeeSongMurphy20}
{Lee} J.,  {Song} I.,   {Murphy} S.,  2020, \mn@doi [\mnras]
  {10.1093/mnras/staa689}, \href
  {https://ui.adsabs.harvard.edu/abs/2020MNRAS.494...62L} {494, 62}

\bibitem[\protect\citeauthoryear{{Manara} et~al.,}{{Manara}
  et~al.}{2016}]{2016A&A...591L...3M}
{Manara} C.~F.,  et~al., 2016, \mn@doi [\aap] {10.1051/0004-6361/201628549},
  \href {https://ui.adsabs.harvard.edu/abs/2016A&A...591L...3M} {591, L3}

\bibitem[\protect\citeauthoryear{{Manara} et~al.,}{{Manara}
  et~al.}{2020}]{2020A&A...639A..58M}
{Manara} C.~F.,  et~al., 2020, \mn@doi [\aap] {10.1051/0004-6361/202037949},
  \href {https://ui.adsabs.harvard.edu/abs/2020A&A...639A..58M} {639, A58}

\bibitem[\protect\citeauthoryear{{Manara} et~al.,}{{Manara}
  et~al.}{2021}]{2021A&A...650A.196M}
{Manara} C.~F.,  et~al., 2021, \mn@doi [\aap] {10.1051/0004-6361/202140639},
  \href {https://ui.adsabs.harvard.edu/abs/2021A&A...650A.196M} {650, A196}

\bibitem[\protect\citeauthoryear{{Mann} et~al.,}{{Mann}
  et~al.}{2014}]{2014ApJ...784...82M}
{Mann} R.~K.,  et~al., 2014, \mn@doi [\apj] {10.1088/0004-637X/784/1/82}, \href
  {https://ui.adsabs.harvard.edu/abs/2014ApJ...784...82M} {784, 82}

\bibitem[\protect\citeauthoryear{{Matsuyama}, {Johnstone}  \&
  {Hartmann}}{{Matsuyama} et~al.}{2003}]{Matsuyama03}
{Matsuyama} I.,  {Johnstone} D.,   {Hartmann} L.,  2003, \mn@doi [\apj]
  {10.1086/344638}, \href
  {https://ui.adsabs.harvard.edu/abs/2003ApJ...582..893M} {582, 893}

\bibitem[\protect\citeauthoryear{{Mordasini}}{{Mordasini}}{2018}]{Mordasini18}
{Mordasini} C.,  2018, in {Deeg} H.~J.,  {Belmonte} J.~A.,  eds, , Handbook of
  Exoplanets.
p.~143, \mn@doi{10.1007/978-3-319-55333-7\_143}

\bibitem[\protect\citeauthoryear{{Nakatani}, {Hosokawa}, {Yoshida}, {Nomura}
  \& {Kuiper}}{{Nakatani} et~al.}{2018}]{2018ApJ...865...75N}
{Nakatani} R.,  {Hosokawa} T.,  {Yoshida} N.,  {Nomura} H.,   {Kuiper} R.,
  2018, \mn@doi [\apj] {10.3847/1538-4357/aad9fd}, \href
  {https://ui.adsabs.harvard.edu/abs/2018ApJ...865...75N} {865, 75}

\bibitem[\protect\citeauthoryear{{{\"O}berg} \& {Bergin}}{{{\"O}berg} \&
  {Bergin}}{2021}]{2021PhR...893....1O}
{{\"O}berg} K.~I.,  {Bergin} E.~A.,  2021, \mn@doi [\physrep]
  {10.1016/j.physrep.2020.09.004}, \href
  {https://ui.adsabs.harvard.edu/abs/2021PhR...893....1O} {893, 1}

\bibitem[\protect\citeauthoryear{{O'dell} \& {Wen}}{{O'dell} \&
  {Wen}}{1994}]{1994ApJ...436..194O}
{O'dell} C.~R.,  {Wen} Z.,  1994, \mn@doi [\apj] {10.1086/174892}, \href
  {https://ui.adsabs.harvard.edu/abs/1994ApJ...436..194O} {436, 194}

\bibitem[\protect\citeauthoryear{{Owen}}{{Owen}}{2016}]{2016PASA...33....5O}
{Owen} J.~E.,  2016, \mn@doi [\pasa] {10.1017/pasa.2016.2}, \href
  {https://ui.adsabs.harvard.edu/abs/2016PASA...33....5O} {33, e005}

\bibitem[\protect\citeauthoryear{{Owen}, {Ercolano}, {Clarke}  \&
  {Alexander}}{{Owen} et~al.}{2010}]{2010MNRAS.401.1415O}
{Owen} J.~E.,  {Ercolano} B.,  {Clarke} C.~J.,   {Alexander} R.~D.,  2010,
  \mn@doi [\mnras] {10.1111/j.1365-2966.2009.15771.x}, \href
  {http://adsabs.harvard.edu/abs/2010MNRAS.401.1415O} {401, 1415}

\bibitem[\protect\citeauthoryear{{Owen}, {Ercolano}  \& {Clarke}}{{Owen}
  et~al.}{2011}]{2011MNRAS.412...13O}
{Owen} J.~E.,  {Ercolano} B.,   {Clarke} C.~J.,  2011, \mn@doi [\mnras]
  {10.1111/j.1365-2966.2010.17818.x}, \href
  {http://adsabs.harvard.edu/abs/2011MNRAS.412...13O} {412, 13}

\bibitem[\protect\citeauthoryear{{Owen}, {Clarke}  \& {Ercolano}}{{Owen}
  et~al.}{2012}]{Owen12}
{Owen} J.~E.,  {Clarke} C.~J.,   {Ercolano} B.,  2012, \mn@doi [\mnras]
  {10.1111/j.1365-2966.2011.20337.x}, \href
  {https://ui.adsabs.harvard.edu/abs/2012MNRAS.422.1880O} {422, 1880}

\bibitem[\protect\citeauthoryear{{Picogna}, {Ercolano}, {Owen}  \&
  {Weber}}{{Picogna} et~al.}{2019}]{Picogna19}
{Picogna} G.,  {Ercolano} B.,  {Owen} J.~E.,   {Weber} M.~L.,  2019, \mn@doi
  [\mnras] {10.1093/mnras/stz1166}, \href
  {https://ui.adsabs.harvard.edu/abs/2019MNRAS.487..691P} {487, 691}

\bibitem[\protect\citeauthoryear{{Picogna}, {Ercolano}  \&
  {Espaillat}}{{Picogna} et~al.}{2021}]{Picogna21}
{Picogna} G.,  {Ercolano} B.,   {Espaillat} C.~C.,  2021, \mn@doi [\mnras]
  {10.1093/mnras/stab2883}, \href
  {https://ui.adsabs.harvard.edu/abs/2021MNRAS.508.3611P} {508, 3611}

\bibitem[\protect\citeauthoryear{{Pinte} et~al.,}{{Pinte}
  et~al.}{2018}]{2018ApJ...860L..13P}
{Pinte} C.,  et~al., 2018, \mn@doi [\apjl] {10.3847/2041-8213/aac6dc}, \href
  {https://ui.adsabs.harvard.edu/abs/2018ApJ...860L..13P} {860, L13}

\bibitem[\protect\citeauthoryear{{Preibisch} et~al.,}{{Preibisch}
  et~al.}{2005}]{2005ApJS..160..401P}
{Preibisch} T.,  et~al., 2005, \mn@doi [\apjs] {10.1086/432891}, \href
  {https://ui.adsabs.harvard.edu/abs/2005ApJS..160..401P} {160, 401}

\bibitem[\protect\citeauthoryear{{Pringle}}{{Pringle}}{1981}]{1981ARA&A..19..137P}
{Pringle} J.~E.,  1981, \mn@doi [\araa] {10.1146/annurev.aa.19.090181.001033},
  \href {https://ui.adsabs.harvard.edu/abs/1981ARA&A..19..137P} {19, 137}

\bibitem[\protect\citeauthoryear{{Pudritz}, {Ouyed}, {Fendt}  \&
  {Brandenburg}}{{Pudritz} et~al.}{2007}]{2007prpl.conf..277P}
{Pudritz} R.~E.,  {Ouyed} R.,  {Fendt} C.,   {Brandenburg} A.,  2007, in
  {Reipurth} B.,  {Jewitt} D.,   {Keil} K.,  eds, Protostars and Planets V.
  p.~277 (\mn@eprint {arXiv} {astro-ph/0603592})

\bibitem[\protect\citeauthoryear{{Qiao}, {Haworth}, {Sellek}  \& {Ali}}{{Qiao}
  et~al.}{2022}]{Qiao22}
{Qiao} L.,  {Haworth} T.~J.,  {Sellek} A.~D.,   {Ali} A.~A.,  2022, \mn@doi
  [\mnras] {10.1093/mnras/stac684}, \href
  {https://ui.adsabs.harvard.edu/abs/2022MNRAS.512.3788Q} {512, 3788}

\bibitem[\protect\citeauthoryear{{Reiter} \& {Parker}}{{Reiter} \&
  {Parker}}{2019}]{2019MNRAS.486.4354R}
{Reiter} M.,  {Parker} R.~J.,  2019, \mn@doi [\mnras] {10.1093/mnras/stz1115},
  \href {https://ui.adsabs.harvard.edu/abs/2019MNRAS.486.4354R} {486, 4354}

\bibitem[\protect\citeauthoryear{{Ribas}, {Bouy}  \& {Mer{\'\i}n}}{{Ribas}
  et~al.}{2015}]{Ribas15}
{Ribas} {\'A}.,  {Bouy} H.,   {Mer{\'\i}n} B.,  2015, \mn@doi [\aap]
  {10.1051/0004-6361/201424846}, \href
  {https://ui.adsabs.harvard.edu/abs/2015A&A...576A..52R} {576, A52}

\bibitem[\protect\citeauthoryear{{Richert}, {Getman}, {Feigelson}, {Kuhn},
  {Broos}, {Povich}, {Bate}  \& {Garmire}}{{Richert} et~al.}{2018}]{Richert18}
{Richert} A.~J.~W.,  {Getman} K.~V.,  {Feigelson} E.~D.,  {Kuhn} M.~A.,
  {Broos} P.~S.,  {Povich} M.~S.,  {Bate} M.~R.,   {Garmire} G.~P.,  2018,
  \mn@doi [\mnras] {10.1093/mnras/sty949}, \href
  {https://ui.adsabs.harvard.edu/abs/2018MNRAS.477.5191R} {477, 5191}

\bibitem[\protect\citeauthoryear{{Richling} \& {Yorke}}{{Richling} \&
  {Yorke}}{2000}]{2000ApJ...539..258R}
{Richling} S.,  {Yorke} H.~W.,  2000, \mn@doi [\apj] {10.1086/309198}, \href
  {https://ui.adsabs.harvard.edu/abs/2000ApJ...539..258R} {539, 258}

\bibitem[\protect\citeauthoryear{{Segura-Cox} et~al.,}{{Segura-Cox}
  et~al.}{2020}]{2020Natur.586..228S}
{Segura-Cox} D.~M.,  et~al., 2020, \mn@doi [\nat] {10.1038/s41586-020-2779-6},
  \href {https://ui.adsabs.harvard.edu/abs/2020Natur.586..228S} {586, 228}

\bibitem[\protect\citeauthoryear{{Sellek}, {Booth}  \& {Clarke}}{{Sellek}
  et~al.}{2020}]{2020MNRAS.492.1279S}
{Sellek} A.~D.,  {Booth} R.~A.,   {Clarke} C.~J.,  2020, \mn@doi [\mnras]
  {10.1093/mnras/stz3528}, \href
  {https://ui.adsabs.harvard.edu/abs/2020MNRAS.492.1279S} {492, 1279}

\bibitem[\protect\citeauthoryear{{Shakura} \& {Sunyaev}}{{Shakura} \&
  {Sunyaev}}{1973}]{Shak}
{Shakura} N.~I.,  {Sunyaev} R.~A.,  1973, \aap, \href
  {http://adsabs.harvard.edu/abs/1973A%26A....24..337S} {24, 337}

\bibitem[\protect\citeauthoryear{{Sicilia-Aguilar} et~al.,}{{Sicilia-Aguilar}
  et~al.}{2016}]{2016PASA...33...59S}
{Sicilia-Aguilar} A.,  et~al., 2016, \mn@doi [\pasa] {10.1017/pasa.2016.56},
  \href {https://ui.adsabs.harvard.edu/abs/2016PASA...33...59S} {33, e059}

\bibitem[\protect\citeauthoryear{{Silverberg} et~al.,}{{Silverberg}
  et~al.}{2016}]{Silverberg16}
{Silverberg} S.~M.,  et~al., 2016, \mn@doi [\apjl]
  {10.3847/2041-8205/830/2/L28}, \href
  {https://ui.adsabs.harvard.edu/abs/2016ApJ...830L..28S} {830, L28}

\bibitem[\protect\citeauthoryear{{Silverberg} et~al.,}{{Silverberg}
  et~al.}{2020}]{Silverberg20}
{Silverberg} S.~M.,  et~al., 2020, \mn@doi [\apj] {10.3847/1538-4357/ab68e6},
  \href {https://ui.adsabs.harvard.edu/abs/2020ApJ...890..106S} {890, 106}

\bibitem[\protect\citeauthoryear{{Smith}, {Bally}, {Shuping}, {Morris}  \&
  {Kassis}}{{Smith} et~al.}{2005}]{2005AJ....130.1763S}
{Smith} N.,  {Bally} J.,  {Shuping} R.~Y.,  {Morris} M.,   {Kassis} M.,  2005,
  \mn@doi [\aj] {10.1086/432912}, \href
  {https://ui.adsabs.harvard.edu/abs/2005AJ....130.1763S} {130, 1763}

\bibitem[\protect\citeauthoryear{{Suzuki} \& {Inutsuka}}{{Suzuki} \&
  {Inutsuka}}{2009}]{2009ApJ...691L..49S}
{Suzuki} T.~K.,  {Inutsuka} S.-i.,  2009, \mn@doi [\apjl]
  {10.1088/0004-637X/691/1/L49}, \href
  {https://ui.adsabs.harvard.edu/abs/2009ApJ...691L..49S} {691, L49}

\bibitem[\protect\citeauthoryear{{Suzuki}, {Ogihara}, {Morbidelli}, {Crida}  \&
  {Guillot}}{{Suzuki} et~al.}{2016}]{2016A&A...596A..74S}
{Suzuki} T.~K.,  {Ogihara} M.,  {Morbidelli} A.,  {Crida} A.,   {Guillot} T.,
  2016, \mn@doi [\aap] {10.1051/0004-6361/201628955}, \href
  {https://ui.adsabs.harvard.edu/abs/2016A&A...596A..74S} {596, A74}

\bibitem[\protect\citeauthoryear{{Tabone}, {Rosotti}, {Lodato}, {Armitage},
  {Cridland}  \& {van Dishoeck}}{{Tabone} et~al.}{2021}]{2021MNRAS.tmpL.109T}
{Tabone} B.,  {Rosotti} G.~P.,  {Lodato} G.,  {Armitage} P.~J.,  {Cridland}
  A.~J.,   {van Dishoeck} E.~F.,  2021, \mn@doi [\mnras]
  {10.1093/mnrasl/slab124}, \href
  {https://ui.adsabs.harvard.edu/abs/2021MNRAS.tmpL.109T} {}

\bibitem[\protect\citeauthoryear{{Teague}, {Bae}  \& {Bergin}}{{Teague}
  et~al.}{2019}]{2019Natur.574..378T}
{Teague} R.,  {Bae} J.,   {Bergin} E.~A.,  2019, \mn@doi [\nat]
  {10.1038/s41586-019-1642-0}, \href
  {https://ui.adsabs.harvard.edu/abs/2019Natur.574..378T} {574, 378}

\bibitem[\protect\citeauthoryear{{Trapman}, {Facchini}, {Hogerheijde}, {van
  Dishoeck}  \& {Bruderer}}{{Trapman} et~al.}{2019}]{2019A&A...629A..79T}
{Trapman} L.,  {Facchini} S.,  {Hogerheijde} M.~R.,  {van Dishoeck} E.~F.,
  {Bruderer} S.,  2019, \mn@doi [\aap] {10.1051/0004-6361/201834723}, \href
  {https://ui.adsabs.harvard.edu/abs/2019A&A...629A..79T} {629, A79}

\bibitem[\protect\citeauthoryear{{Turner}, {Fromang}, {Gammie}, {Klahr},
  {Lesur}, {Wardle}  \& {Bai}}{{Turner} et~al.}{2014}]{2014prpl.conf..411T}
{Turner} N.~J.,  {Fromang} S.,  {Gammie} C.,  {Klahr} H.,  {Lesur} G.,
  {Wardle} M.,   {Bai} X.~N.,  2014, in {Beuther} H.,  {Klessen} R.~S.,
  {Dullemond} C.~P.,   {Henning} T.,  eds, Protostars and Planets VI. p.~411
  (\mn@eprint {arXiv} {1401.7306}),
  \mn@doi{10.2458/azu\_uapress\_9780816531240-ch018}

\bibitem[\protect\citeauthoryear{{Wang} \& {Goodman}}{{Wang} \&
  {Goodman}}{2017}]{2017ApJ...847...11W}
{Wang} L.,  {Goodman} J.,  2017, \mn@doi [\apj] {10.3847/1538-4357/aa8726},
  \href {https://ui.adsabs.harvard.edu/abs/2017ApJ...847...11W} {847, 11}

\bibitem[\protect\citeauthoryear{{Wilhelm} \& {Portegies Zwart}}{{Wilhelm} \&
  {Portegies Zwart}}{2022}]{wilhelm22}
{Wilhelm} M. J.~C.,  {Portegies Zwart} S.,  2022, \mn@doi [\mnras]
  {10.1093/mnras/stab2523}, \href
  {https://ui.adsabs.harvard.edu/abs/2022MNRAS.509...44W} {509, 44}

\bibitem[\protect\citeauthoryear{{Winter}, {Clarke}, {Rosotti}, {Hacar}  \&
  {Alexander}}{{Winter} et~al.}{2019}]{2019MNRAS.490.5478W}
{Winter} A.~J.,  {Clarke} C.~J.,  {Rosotti} G.~P.,  {Hacar} A.,   {Alexander}
  R.,  2019, \mn@doi [\mnras] {10.1093/mnras/stz2545}, \href
  {https://ui.adsabs.harvard.edu/abs/2019MNRAS.490.5478W} {490, 5478}

\bibitem[\protect\citeauthoryear{{Winter}, {Kruijssen}, {Chevance}, {Keller}
  \& {Longmore}}{{Winter} et~al.}{2020}]{2020MNRAS.491..903W}
{Winter} A.~J.,  {Kruijssen} J.~M.~D.,  {Chevance} M.,  {Keller} B.~W.,
  {Longmore} S.~N.,  2020, \mn@doi [\mnras] {10.1093/mnras/stz2747}, \href
  {https://ui.adsabs.harvard.edu/abs/2020MNRAS.491..903W} {491, 903}

\bibitem[\protect\citeauthoryear{{Wyatt}}{{Wyatt}}{2008}]{Wyatt08}
{Wyatt} M.~C.,  2008, \mn@doi [\araa] {10.1146/annurev.astro.45.051806.110525},
  \href {http://adsabs.harvard.edu/abs/2008ARA%26A..46..339W} {46, 339}

\bibitem[\protect\citeauthoryear{{Yu}, {Evans}, {Dodson-Robinson}, {Willacy}
  \& {Turner}}{{Yu} et~al.}{2017}]{2017ApJ...841...39Y}
{Yu} M.,  {Evans} Neal~J. I.,  {Dodson-Robinson} S.~E.,  {Willacy} K.,
  {Turner} N.~J.,  2017, \mn@doi [\apj] {10.3847/1538-4357/aa6e4c}, \href
  {https://ui.adsabs.harvard.edu/abs/2017ApJ...841...39Y} {841, 39}

\makeatother
\end{thebibliography}

\appendix
\section{List of disc evolution pathways}
\begin{table*}
    \centering
    \begin{tabularx}{\linewidth}{|c|L|L|L|} 
    \hline
    Acronym & Dispersal Pathway & Criteria & Description \\
    \hline
        LL & Long-lived discs & $\tau_{\rm disc}>20$ Myr & Discs with lifetimes longer than 20 Myr. Requires low internal and external photoevaporation rates, allowing the discs to slowly accrete on to the central stars. \\
    \hline
        IO & Inside-Out \newline\newline Panel 4 of fig. \ref{fig:evolution_cartoon}& $\tau_{\rm hole} < \tau_{\rm disc}$ & Internal photoevaporation and accretion open up an inner hole, resulting in an enhanced internal photoevaporative wind being driven off of the inner edge of the disc. The enhanced wind causes the disc to disperse from the inside-out with the hole growing in size over time until the disc undergoes full disc dispersal.\\
    \hline
        IO-A & Inside-Out with continued accretion \newline\newline Panel 3 of fig. \ref{fig:evolution_cartoon}& 1) $\tau_{\rm gap} < \tau_{\rm disc} < \tau_{\rm hole}$ \newline\newline 2) $r_{\rm outer}>r_{\rm clear}$ & Similar pathway to inside out but with an inner accreting disc. Internal photoevaporation dominates the evolution of the disc in the middle and outer regions, but not the very inner disc. This leads to a gap forming in the disc and the outer disc receding past the clearing radius due to internal photoevaporation. Eventually the outer disc is fully dispersed by the internal irradiation leaving only an inner disc that slowly accretes on to the central star.\\
    \hline
        OI & Outside-In \newline\newline Panel 1 of fig. \ref{fig:evolution_cartoon}& $\tau_{\rm disc} > \tau_{\rm gap}$ & External photoevaporation dominates the evolution, truncating the disc over time. As external photoevaporation becomes ineffective close to the central star, the truncated disc accretes on to the central star whilst resupplying more modest levels of external photoevaporation through viscous spreading. This continues with the disc gradually reducing in size and mass until the last remnants have accreted on to the central star.\\
    \hline
        OI-G & Outside-In with gap features \newline\newline Panel 2 of fig. \ref{fig:evolution_cartoon} & 1) $\tau_{\rm gap} < \tau_{\rm disc} < \tau_{\rm hole}$ \newline\newline 2) $r_{\rm outer}<r_{\rm clear}$& The initial pathway is similar to outside-in, but with the difference being that the internal photoevaporation rate is sufficient to open a gap in the disc before external photoevaporation has been able to truncate the disc down to the effective gravitational radius. As the disc evolves, external photoevaporation truncates the outer disc region past the clearing radius before internal irradiation can remove disc. Once the disc is fully dispersed, an accreting inner disc remains.\\
    \hline
    \end{tabularx}
    \caption{Types of final disc evolution pathways, with their acronyms and descriptions}
    \label{tab:dispersal_types}
\end{table*}

\vspace{-0.2cm}
\label{lastpage}
\end{document}